\documentclass[trackchanges]{aastex701}
\usepackage{placeins}
\usepackage{graphicx}
\usepackage{color}
\usepackage{amsmath}
\graphicspath{{./}{figures/}}

\begin{document}

\title{Physics-Informed Neural Networks for Fast Multilayer Spectral Inversion of \\H$\alpha$ 6562.8~\AA\ and Ca~II 8542.1~\AA\ Spectra}

\author{Ziyang Zhang}
\email{zz57@njit.edu}
\affiliation{Department of Mechanical and Industrial Engineering, New Jersey Institute of Technology, Newark, NJ, USA}

\author{Qin Li}
\email{qin.li@njit.edu}
\affiliation{Department of Physics, New Jersey Institute of Technology, Newark, NJ, USA}

\author{Vasyl B. Yurchyshyn}
\email{vasyl.yurchyshyn@njit.edu}
\affiliation{Department of Physics, New Jersey Institute of Technology, Newark, NJ, USA}
\affiliation{Big Bear Solar Observatory, New Jersey Institute of Technology, Big Bear City, CA, USA}

\author{Kangwoo Yi}
\email{kangwoo.yi@njit.edu}
\affiliation{Department of Physics, New Jersey Institute of Technology, Newark, NJ, USA}

\author{Haimin Wang}
\email{haimin.wang@njit.edu}
\affiliation{Department of Physics, New Jersey Institute of Technology, Newark, NJ, USA}
\affiliation{Big Bear Solar Observatory, New Jersey Institute of Technology, Big Bear City, CA, USA}

\author{Wenda Cao}
\email{wenda.cao@njit.edu}
\affiliation{Department of Physics, New Jersey Institute of Technology, Newark, NJ, USA}
\affiliation{Big Bear Solar Observatory, New Jersey Institute of Technology, Big Bear City, CA, USA}

\author{Bo Shen}
\email{bo.shen@njit.edu}
\affiliation{Department of Mechanical and Industrial Engineering, New Jersey Institute of Technology, Newark, NJ, USA}
\affiliation{Department of Data Science, New Jersey Institute of Technology, Newark, NJ, USA}
\correspondingauthor{Bo Shen}
\email[show]{bo.shen@njit.edu}

\begin{abstract}
Strong chromospheric absorption lines such as H$\alpha$ 6562.8~\AA\ and Ca~II 8542.1~\AA\ provide important diagnostics of plasma dynamics and thermal structure in the solar chromosphere. Multilayer spectral inversion (MLSI) offers a physically interpretable framework for modeling these lines with a finite number of radiative-transfer layers, but conventional MLSI still requires pixel-by-pixel nonlinear least-squares fitting and is therefore costly for large imaging spectroscopic data sets. In this work, we introduce a physics-informed neural-network (PINN) framework to accelerate MLSI while retaining its analytic radiative-transfer formulation. The network predicts MLSI parameters from observed line profiles, then passes the predicted parameters through the differentiable MLSI forward model to synthesize spectra. We train the model in two stages: the first stage uses only the spectral reconstruction loss, followed by a second fine-tuning stage that combines spectral consistency with parameter-space supervision from conventional MLSI results for a single reference image. This strategy reduces the need for large precomputed training sets while preserving the physical interpretability of the MLSI parameters. We apply the method to FISS H$\alpha$ 6562.8~\AA\ and Ca~II 8542.1~\AA\ observations of both quiet-Sun and active-region targets. The MLSI-PINN parameter maps reproduce the main spatial structures of direct MLSI inversions, taken with the Fast Imaging Solar Spectrograph (FISS) of the Goode Solar Telescope (GST). Across all parameter comparisons shown for the representative quiet-Sun and active-region rasters, the arithmetic mean of the pixel-wise Pearson correlation coefficients is 0.933. The reconstructed spectra also remain close to both the observed profiles and conventional MLSI reconstructions. With the implementations and hardware used in this study, MLSI-PINN requires approximately 5--15~s to process one raster after training, compared with approximately 3--5~min for conventional MLSI, corresponding to an inference speedup of about 12--60 times. These results show that the two-stage physics-informed neural network provides a substantially faster, physically constrained approximation to MLSI without a dominant loss in reconstruction quality, improving the feasibility of applying MLSI to large chromospheric imaging-spectroscopy data sets.
\end{abstract}

\section{Introduction}

High-resolution spectroscopy of chromospheric lines provides diagnostics of the thermal and dynamic structure of the lower solar atmosphere through analysis of line profiles. Among the commonly used diagnostics, H$\alpha$ and Ca~II 8542.1~\AA\ lines are particularly important because their profiles sample atmospheric conditions from the upper photosphere into the chromosphere and carry complementary information on opacity, line-of-sight motions, and line broadening \citep{Leenaarts2009,Leenaarts2012,Carlsson2019}. The Fast Imaging Solar Spectrograph (FISS) at the Goode Solar Telescope provides a complete spectral profile over a defined field of view \citep{Goode2012,Chae2013,Chae2021}. Such observations contain substantially more information than monochromatic images, but quantitative analysis of the atmospheric parameters requires an appropriate inversion model.
 
Inversions of H$\alpha$ and Ca~II spectra have been used to investigate the temperature and velocity structure of umbral flashes \citep{Henriques2017,Felipe2019}. Related analyses have constrained the thermal, velocity, and line-broadening properties of chromospheric fibrils and transient brightenings \citep{Kianfar2020,Vissers2019}. With full-Stokes Ca~II observations, spectropolarimetric inversions can also constrain magnetic fields in plage \citep{Pietrow2020}. Multiline spectropolarimetric inversions of DKIST/ViSP observations have also been used to diagnose the thermal and dynamic structure of flare ribbons \citep{Yadav2025}.  These applications show that the shape of the spectral line carries thermal and dynamic information in addition to morphology alone. The detailed shape of the spectral line reflects the combined effects of temperature, velocity, opacity, and radiative transfer, which also provides constraints on the atmospheric state. At the same time, chromospheric line formation is sufficiently complex so that extracting this information is rarely a simple measurement problem. More physically detailed inversion models generally require more expensive radiative-transfer calculations and nonlinear optimization, which can become a limiting factor for large spectroscopic datasets.

These computational demands motivate a practical compromise between the physical detail of the adopted atmospheric model and computational feasibility. Compact analytic approaches, including Milne--Eddington formulations \citep{Unno1956,Skumanich1987} and cloud-model inversions \citep{Beckers1964,Tziotziou2007}, represent the atmosphere with a small number of parameters and can be applied efficiently to large data volumes. Such models have also been used to analyze chromospheric structures \citep{Chae2014}. Their efficiency, however, comes from restrictive assumptions about the atmospheric structure. More flexible stratified inversion codes, including SIR and NICOLE \citep{RuizCobo1992,SocasNavarro2015} and STiC and SNAPI \citep{deLaCruzRodriguez2019,Milic2018}, can retrieve height-dependent atmospheric quantities. Similar non-local thermodynamic equilibrium methods have also been applied to Mg~II spectra \citep{deLaCruzRodriguez2016}. All iterative inversion methods require repeated forward-model evaluations; the main differences in computational cost arise from the complexity of the forward model and the number of free parameters. \citep{RuizCobo1992,SocasNavarro2015,deLaCruzRodriguez2016,deLaCruzRodriguez2019,Milic2018}. For high-cadence rasters and long observing sequences, this repeated pixel-by-pixel fitting can dominate the analysis cost.

Multilayer spectral inversion (MLSI) was developed to model strong absorption lines that form over a broad height range from the photosphere to the chromosphere. It combines a Milne--Eddington description of the photospheric contribution with a cloud-model description of the chromospheric contribution within a finite number of radiative-transfer layers \citep{Chae2020}. The method returns parameters with explicit physical meanings, including source functions, Doppler velocities, Doppler widths, opacity-related quantities, damping, and total chromospheric optical thickness. A subsequent extension allowed the chromospheric absorption profile to vary with height \citep{Chae2021}. For H$\alpha$ and Ca~II~8542, this formulation retains an analytic forward model while allowing more vertical structure than a single-layer cloud model. Nevertheless, conventional MLSI still relies on constrained nonlinear least-squares fitting at each pixel. Thus, although MLSI is computationally lighter than stratified inversion under non-local thermodynamic equilibrium, applying it repeatedly to many FISS rasters or long time sequences remains costly. 

Machine-learning methods provide a possible route to accelerate this type of inverse problem. Once trained, a neural network can replace repeated iterative fitting with a direct evaluation of the inverse mapping. Neural networks have been used to approximate spectroscopic and spectropolarimetric inversion mappings, and machine-learning methods are increasingly used in solar diagnostic applications more broadly \citep{AsensioRamos2019,SainzDalda2019,Osborne2019,AsensioRamos2023}. The SPIn4D project provides radiative-MHD simulations and synthetic Stokes profiles to support the development of deep-learning spectropolarimetric inversions \citep{Yang2024SPIn4D}. Fast numerical approaches have also been developed for related inverse problems, such as differential emission measure reconstruction \citep{Cheung2015}. In the MLSI context, \citet{Lee2022} showed that a supervised neural network can reproduce MLSI outputs for H$\alpha$ and Ca~II~8542 much faster than direct nonlinear fitting. Although its inference is fast, the model is trained exclusively against precomputed MLSI parameter labels and therefore requires a sufficiently large and representative set of conventional inversion results. Its performance consequently depends on the coverage and consistency of the training labels, and ambiguities or systematic biases in the reference inversions may be inherited by the network.

Physics-informed neural networks (PINN) provide a way to include the forward model more directly in the learning process \citep{Raissi2019,Karniadakis2021,gnanasambandam2023self}. Instead of training only against parameter labels, the predicted parameters can be evaluated through the physical model that maps them back to observables. In solar physics, related ideas have been applied to coronal magnetic-field modeling, neural-field-based spectropolarimetric inference, and Milne--Eddington inversion \citep{Jarolim2023NatAs,DiazBaso2025,Jarolim2025PINNME,li2025mvpinn}. Physics-informed learning has also been used to reconstruct three-dimensional magnetic fields from spectropolarimetric inversion results \citep{Yang2025SPIn4D}, whereas our work focuses on intensity-only spectroscopic inversion. MLSI is a natural candidate for this strategy because its forward calculation is analytic and differentiable. The MLSI parameters predicted by a neural network can be passed through the multilayer radiative-transfer model during training, so that the loss is defined at least in part by the mismatch between observed and synthesized spectra.

In this work, we develop an MLSI-PINN framework for fast inversion of FISS H$\alpha$ and Ca~II~8542 spectra taken with the Fast Imaging Solar Spectrograph (FISS) of Goode Solar Telescope (GST). The network takes an observed spectral profile as input and predicts an MLSI parameter vector. The embedded MLSI forward model then synthesizes the corresponding emergent line profile, and the spectral reconstruction error provides the primary physics-based constraint. To reduce degeneracy among multilayer solutions, we use a second training stage in which a limited conventional MLSI reference raster supplies parameter-space supervision for selected quantities. This design is intended to preserve the interpretability of MLSI parameters while reducing the need for repeated pixel-by-pixel nonlinear optimization. The method should therefore be viewed not as a replacement for detailed radiative-transfer inversions, but as an acceleration strategy for MLSI-style analysis of large FISS data sets and time-dependent observations.

The remainder of this paper is organized as follows. Section~\ref{sec:method} describes the MLSI forward model and the two-stage MLSI-PINN training framework. Section~\ref{sec:data} presents the FISS observations and preprocessing procedures. Section~\ref{sec:results} evaluates spectral reconstruction, parameter-space agreement, temporal behavior, and computational performance. Section~\ref{sec:discussion} summarizes the implications and limitations of the proposed approach.

\section{Method}
\label{sec:method}

\subsection{MLSI Forward Model}
\label{subsec:mlsi_forward}

The analytic forward model of multilayer spectral inversion (MLSI) provides the physical component of our method. In the MLSI-PINN framework, a neural network maps an observed spectral profile to a set of MLSI parameters, which are then passed through the analytic radiative-transfer operator to produce a synthetic spectrum. Direct comparison between the synthesized and observed profiles provides the physics-based training signal. In this way, the network can be optimized efficiently while the inferred quantities retain the physical interpretation of the original MLSI parameters.

\begin{figure*}[t]
\centering
\includegraphics[width=\textwidth]{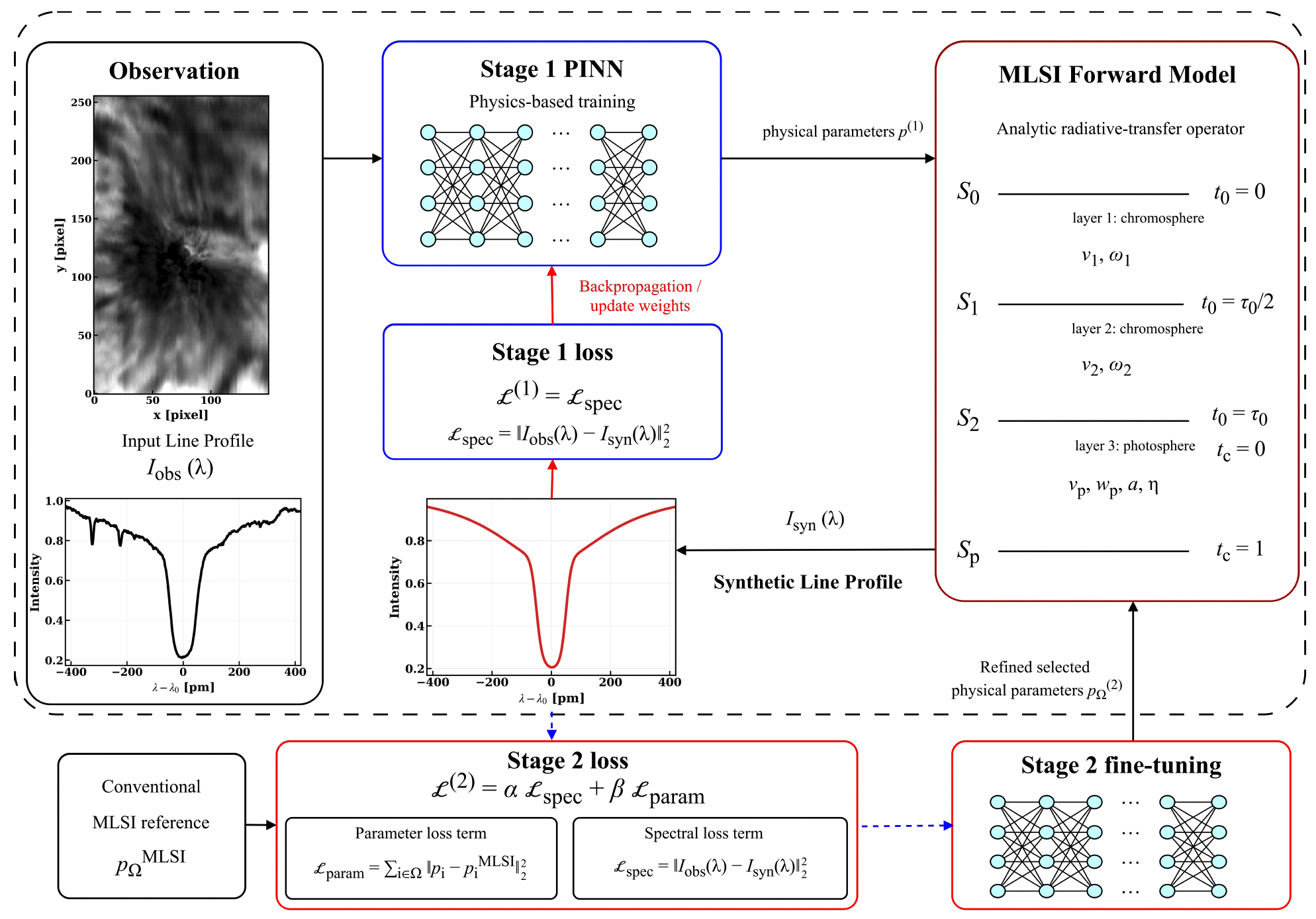}
\caption{Schematic diagram of the two-stage MLSI-PINN framework. In stage 1, the observed spectral profile is mapped to MLSI parameters by a physics-informed neural network and passed through the analytic MLSI forward model to synthesize a spectrum. The network is trained using the spectral reconstruction loss between the observed and synthesized profiles. In stage 2, the stage-1 checkpoint is fine-tuned using a combined loss that includes both the spectral reconstruction loss and a parameter-space loss computed from selected conventional MLSI reference parameters.}
\label{fig:workflow}
\end{figure*}

To establish the notation used below, we briefly summarize the MLSI forward calculation adopted in this work, following the multilayer formulation of \citet{Chae2020,Chae2021}. The calculation begins with the formal solution of the radiative-transfer equation,
\begin{equation}
I_\lambda
=
\int_0^\infty
S(t_\lambda)\,e^{-\tau_\lambda}\,d\tau_\lambda ,
\end{equation}
where $I_\lambda$ is the emergent specific intensity, $S(\cdot)$ is the source function, and $\tau_\lambda$ is the optical depth at wavelength $\lambda$. MLSI approximates this continuous atmosphere with a small number of layers, allowing the emergent spectrum to be expressed analytically in terms of a finite set of physically interpretable parameters.

The atmosphere is represented by a photospheric layer and two chromospheric layers. The photospheric background profile is described by a Voigt-like absorption profile,
\begin{equation}
r(\lambda)
=
1+\eta\,\mathcal{V}
\bigl(\lambda;\Delta\lambda(v_p),w_p,a\bigr),
\end{equation}
which gives the background intensity
\begin{equation}
I_2(\lambda)
=
S_2+\frac{S_p-S_2}{r(\lambda)}.
\end{equation}
Here $r(\lambda)$ is the total-to-continuum absorption factor, $\mathcal{V}(\cdot)$ is the normalized Voigt profile, and $\Delta\lambda(v_p)$ is the wavelength shift associated with the photospheric line-of-sight velocity $v_p$. The quantities $\eta$, $w_p$, and $a$ denote the line-to-continuum opacity ratio, the photospheric Doppler width, and the dimensionless damping parameter, respectively. The continuum optical-depth coordinate is denoted by $\tau_c$, with $S_2$ and $S_p$ representing the source functions at $\tau_c=0$ and $\tau_c=1$, respectively.

The two chromospheric layers are subsequently applied to this photospheric background. Let $t_0$ denote the total chromospheric optical thickness at line center. The source functions $S_0$, $S_1$, and $S_2$ are specified at the line-center optical-depth positions $0$, $\tau_0/2$, and $\tau_0$, respectively. The upper chromospheric layer is characterized by the line-of-sight velocity $v_1$ and Doppler width $\omega_1$, whereas the lower chromospheric layer is characterized by $v_2$ and $\omega_2$. Gaussian-like absorption profiles are adopted for both chromospheric layers. 

The complete 13-dimensional MLSI parameter vector is written as
\begin{equation}
\boldsymbol{p}
=
\left(
v_p,\log\eta,\log w_p,\log a,\log S_p,\log S_2,
\log \tau_0,v_2,v_1,\log w_2,\log w_1,\log S_1,\log S_0
\right).
\end{equation}
Using this parameter vector, the final emergent spectrum can be expressed compactly as
\begin{equation}
I_0(\lambda)
=
\mathcal{F}_{\mathrm{MLSI}}(\boldsymbol{p}),
\end{equation}
where $\mathcal{F}_{\mathrm{MLSI}}(\cdot)$ denotes the complete deterministic MLSI forward operator. Herein, the photospheric background $I_2(\lambda)$ is first propagated through the lower chromospheric layer to obtain the intermediate profile $I_1(\lambda)$ and then through the upper chromospheric layer to obtain the final emergent profile $I_0(\lambda)$.

\subsection{Physics-Informed Neural Network Framework}
\label{subsec:pinn}

As illustrated in Figure~\ref{fig:workflow}, we formulate the acceleration of MLSI as a physics-informed inverse problem. Instead of solving an independent nonlinear least-squares problem for every spatial pixel, a neural network learns an approximate mapping from observed spectral profiles to MLSI parameters. The analytic MLSI forward model remains embedded in the training procedure, allowing the predicted parameters to be evaluated through the spectra they produce rather than solely through an empirical regression target. Once training is complete, the model can infer the MLSI parameters of an observed profile with a single network forward pass.

We denote the neural network by $\mathcal{N}_{\theta}$, where $\theta$ represents its trainable weights. Given a normalized observed spectral profile $I_{\mathrm{obs}}(\lambda)$, the stage-1 network predicts the complete MLSI parameter vector,
\begin{equation}
\boldsymbol{p}^{(1)}
=
\mathcal{N}_{\theta}
\bigl(I_{\mathrm{obs}}(\lambda)\bigr).
\end{equation}
 Before entering the MLSI forward operator, the predicted parameters are mapped to their physically allowed ranges. Positive quantities, including optical thicknesses, Doppler widths, and opacity ratios, are represented in logarithmic form, while bounded quantities are restricted using scaled nonlinear transformations. These constraints prevent the network from exploring nonphysical regions of parameter space and preserve the interpretation of its outputs as MLSI atmospheric parameters.

The spectrum corresponding to the predicted parameter vector is synthesized using the analytic MLSI forward model,
\begin{equation}
I_{\mathrm{syn}}(\lambda)
=
\mathcal{F}_{\mathrm{MLSI}}
\bigl(\boldsymbol{p}^{(1)}\bigr),
\end{equation}
where $I_{\mathrm{syn}}(\lambda)$ is the synthesized emergent profile. The spectral reconstruction loss is then defined as
\begin{equation}
\mathcal{L}_{\mathrm{spec}}
=
\frac{1}{N_\lambda}
\sum_{\lambda}
\left[
I_{\mathrm{syn}}(\lambda)
-
I_{\mathrm{obs}}(\lambda)
\right]^2,
\end{equation}
where $N_\lambda$ is the number of wavelength samples included in the loss. In practice, the loss is evaluated over a central wavelength range around the line core because the far-wing and continuum samples can contain isolated spikes or residual calibration errors that are not adequately represented by the MLSI line-formation model. The selected wavelength range contains the strongest chromospheric contribution and therefore provides the most relevant constraint on the MLSI parameters.

The spectral reconstruction loss alone does not always determine every MLSI parameter uniquely, since different parameter combinations may produce similar emergent profiles. To reduce this ambiguity, we train the model in two stages. During the first stage, the network is optimized using only the physics-based spectral reconstruction loss,
\begin{equation}
\mathcal{L}^{(1)}
=
\mathcal{L}_{\mathrm{spec}}.
\end{equation}
This stage establishes a physics-informed baseline mapping from the observed spectra to the MLSI parameter space without requiring a large set of precomputed MLSI reference parameters.

In the second stage, the stage-1 model provides the initialization, and its objective is augmented with a parameter-space supervision term derived from conventional MLSI inversions of a single reference raster. The refinement focuses on parameters for which the stage-1 predictions exhibit systematic deviations from the conventional inversion. Let $\Omega$ denote the selected subset of MLSI parameters. The parameter loss is evaluated only over this subset,
\begin{equation}
\mathcal{L}_{\mathrm{param}}
=
\frac{1}{|\Omega|}
\sum_{i\in\Omega}
\left(
\boldsymbol{p}_i-\boldsymbol{p}_i^{\mathrm{MLSI}}
\right)^2,
\end{equation}
where $\boldsymbol{p}_i$ is the stage-2 prediction of parameter $i$, $p_i^{\mathrm{MLSI}}$ is the corresponding conventional MLSI reference value, and $|\Omega|$ is the number of selected parameters. The predicted and reference subvectors are denoted by $\boldsymbol{p}_{\Omega}^{(2)}$ and $\boldsymbol{p}_{\Omega}^{\mathrm{MLSI}}$, respectively.

The complete stage-2 objective is
\begin{equation}
\mathcal{L}^{(2)}
=
\alpha\mathcal{L}_{\mathrm{spec}}
+
\beta\mathcal{L}_{\mathrm{param}},
\end{equation}
where $\alpha$ and $\beta$ control the relative contributions of the spectral and parameter-space constraints.

This two-stage procedure retains the spectral constraint imposed by the analytic forward model while using a limited amount of conventional MLSI output to refine selected parameters. As a representative implementation, the neural network used for the H$\alpha$ active-region data is a fully connected multilayer perceptron with hidden layers containing 256, 128, and 64 neurons. Batch normalization and SiLU activations are used in the hidden layers, and a Tanh activation is applied after the third hidden layer. The final linear layer returns the raw MLSI parameter vector. The Ca~II~8542 and quiet-Sun models follow the same general architecture and training procedure, with minor adjustments to the loss windows and parameter ranges to account for differences in spectral sampling and line properties.

For this representative H$\alpha$ setting, stage 1 is trained for 50 epochs using the Adam optimizer with a learning rate of $2\times10^{-4}$, a batch size of 256, a weight decay of $10^{-5}$, and a dropout rate of 0.0. The stage-1 spectral reconstruction loss is evaluated over wavelength indices 180--490. Stage 2 is initialized from the stage-1 checkpoint and fine-tuned for 20 epochs using a learning rate of $2\times10^{-5}$ and zero weight decay. During this stage, the spectral loss is evaluated over a narrower line-core window corresponding to indices 235--345 and is combined with the parameter-space loss using equal weights. The refinement subset is
\begin{equation}
\Omega
=
\{
\log S_1,\log S_0,v_1,v_2,\log w_1,\log w_2
\}.
\end{equation}
Parameters outside $\Omega$ retain their stage-1 values, whereas those within $\Omega$ are replaced by the refined predictions $\boldsymbol{p}_{\Omega}^{(2)}$. The photospheric velocity $v_p$ is handled separately: it is estimated from the observed profile using the conventional line-center proxy and inserted into the final parameter vector rather than being freely refined by the network. The assembled parameter vector is then passed through $\mathcal{F}_{\mathrm{MLSI}}$ to produce the reconstructed spectrum.

Training is performed with mini-batches of spectral profiles sampled from the reference raster, without a validation split or early stopping. After training, the model can be applied to additional observations of the same spectral line obtained on the same observing day. Inference requires only a neural-network forward pass, followed by the analytic MLSI forward calculation when reconstructed spectra are required. This replaces repeated pixel-by-pixel nonlinear fitting with a fixed trained inverse mapping while preserving the physical interpretation of the MLSI parameters.

\FloatBarrier
\section{Data}
\label{sec:data}

\subsection{FISS Observations}
\label{subsec:fiss_obs}

The observational data used in this study were obtained with the Fast Imaging Solar Spectrograph FISS \citep{Chae2013} installed on the 1.6~m Goode Solar Telescope at Big Bear Solar Observatory (BBSO). FISS is a dual-channel imaging spectrograph that simultaneously records the H$\alpha$ and Ca~II~8542~\AA\ spectral bands. For the observing configurations used in this study, the spatial sampling is approximately $0.16''$ per pixel, and the raster cadence ranged from approximately 20 to 35~s.

We analyze two FISS observing sequences. The first sequence is a quiet-Sun observation obtained on 2021 August~7 from 16:40 to 17:23~UT. It contains 105 image frames, and the processed spectral cubes used in this work have dimensions of $502\times246\times200$, where the three axes correspond to wavelength, slit position, and scan position, respectively. The second sequence is an active-region observation obtained on 2023 August~14 from 17:25 to 18:08~UT. It contains 83 image frames, with processed cube dimensions of $512\times256\times150$. Both sequences include simultaneous H$\alpha$ and Ca~II~8542 spectra.

The spectral coverage is approximately 9.7~\AA\ for H$\alpha$ and 12.9~\AA\ for Ca~II~8542, with spectral samplings of about 0.019~\AA\ and 0.026~\AA, respectively. Each image therefore contains on the order of $10^4$--$10^5$ spatially resolved line profiles. These data provide both quiet-Sun and active-region examples for evaluating the proposed MLSI-PINN framework. A summary of the observing sequences is given in Table~\ref{tab:fiss_data}.

\begin{table}[!htbp]
\centering
\caption{Summary of the FISS observing sequences used in this study.}
\label{tab:fiss_data}
\begin{tabular}{cccccc}
\hline
Obs. & Target & Date & UT & Frames & Cube Size \\
\hline
1 & Quiet Sun & 2021 Aug 7 & 16:40--17:23 & 105 & $502\times246\times200$ \\
2 & Active Region & 2023 Aug 14 & 17:25--18:08 & 83 & $512\times256\times150$ \\
\hline
\end{tabular}
\begin{flushleft}
\footnotesize
Note. Both observing sequences include simultaneous H$\alpha$ and Ca~II~8542 spectra. Cube size is given as $N_\lambda\times N_y\times N_x$.
\end{flushleft}\vspace{-0.2in}
\end{table}

\subsection{Data Reduction and Preprocessing}
\label{subsec:data_pre}

The raw FISS data are reduced with the standard FISS reduction pipeline implemented in the \texttt{FISSPy} package, following the procedures described by \citet{Chae2013}. The reduction first removes detector signatures through dark and bias subtraction and flat-field correction. It then corrects the spectrogram geometry, including the spectral tilt and the geometric distortion between the dispersion and slit directions, so that the wavelength and spatial axes are placed on a rectified grid. Wavelength calibration is applied to assign a physical wavelength scale to each spectral channel. The output of this procedure is a calibrated spectral cube for each image frame and for each spectral line.

Because H$\alpha$ and Ca~II~8542 have different wavelength coverages and spectral samplings, the two lines are processed and modeled separately. For each line, the reduced spectra are represented on a fixed wavelength grid before being supplied to the neural network. This keeps the network input dimension fixed and ensures that the MLSI forward operator is evaluated on the same wavelength samples used by the observations. Only the intensity profile is used as the network input in this work.

For each observing sequence and spectral line, one reference image is selected for training. The spectra from this image are used to train the first-stage physics-informed model. The same reference image is also inverted with the conventional MLSI procedure, and the resulting MLSI parameter maps are used as parameter-space supervision targets in the second training stage. The remaining image frames are not used to optimize the network; they are reserved for inference and evaluation after the model has been trained.

Before training, each spectral profile is normalized by its continuum intensity,
\begin{equation}
\tilde{I}(\lambda) =
\frac{I(\lambda)}
{\langle I(\lambda)\rangle_{\mathrm{continuum}}},
\end{equation}
where $\langle I(\lambda)\rangle_{\mathrm{continuum}}$ is the mean intensity over selected line-continuum wavelength samples. The same normalization is applied to the spectra used in the reconstruction loss and during inference. This profile-wise normalization brings the input intensities to a comparable scale, typically close to unity in the line continuums, and improves the numerical conditioning of neural-network training.

\section{Results}
\label{sec:results}

In this section, we evaluate the MLSI-PINN results using the FISS H$\alpha$ and Ca~II~8542 observations described in Section~\ref{sec:data}. The conventional MLSI inversion is used as the reference solution for parameter-space comparison, while the observed spectra provide the reference for spectral-profile reconstruction. We first describe the spectral morphology of the quiet-Sun and active-region observations, then compare the inferred MLSI parameters in map space and pixel-by-pixel parameter space. We subsequently examine representative spectral reconstructions and temporal evolution, and finally summarize the computational performance of the proposed framework.

\begin{figure}[!htbp]
\centering
\includegraphics[width=\textwidth]{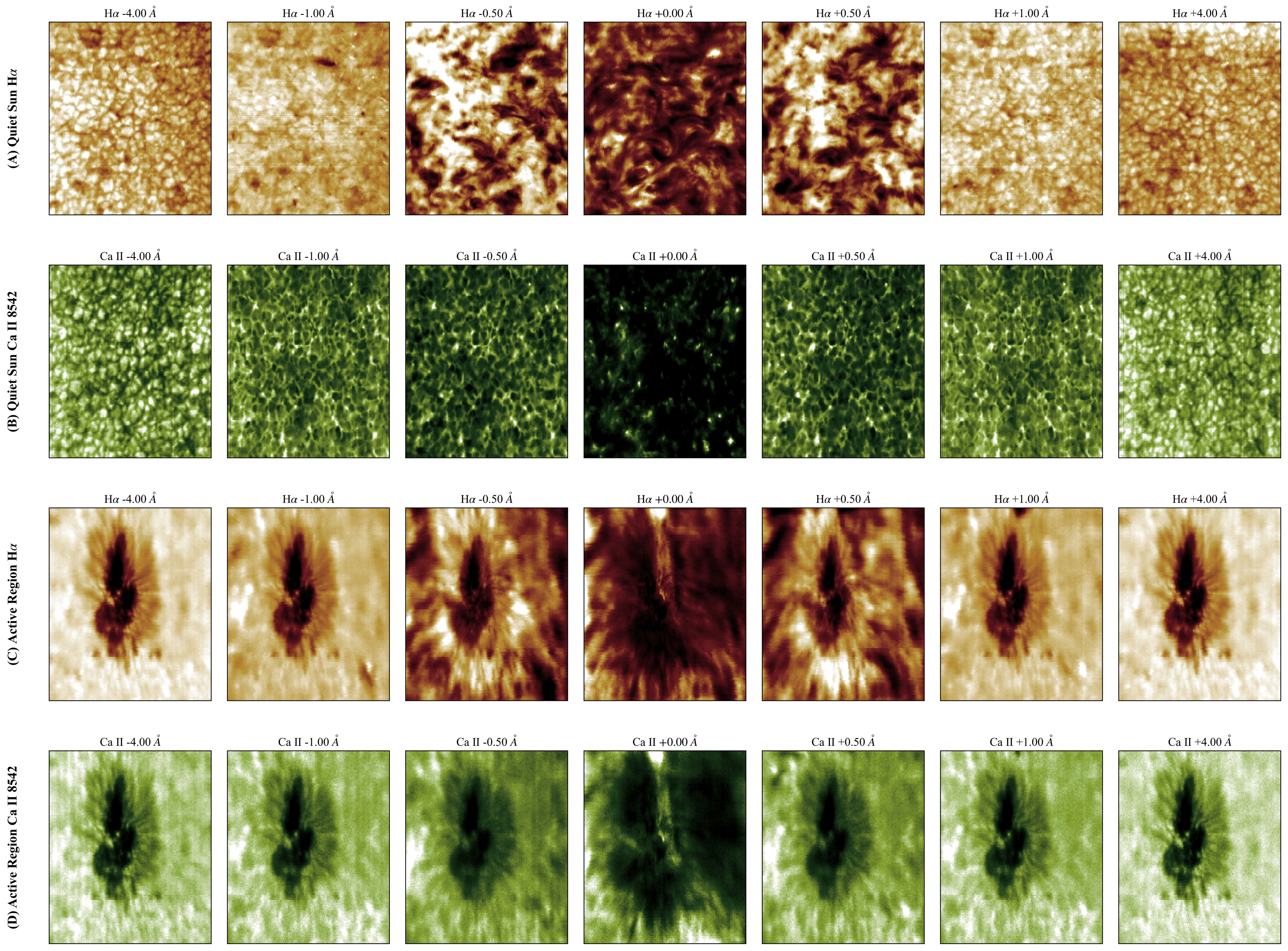}
\caption{Representative quiet-Sun and active-region FISS monochromatic images. Rows (A) and (B) show quiet-Sun H$\alpha$ and Ca~II~8542, respectively, while rows (C) and (D) show the corresponding active-region observations. Each row samples wavelength offsets of $-4$, $-1$, $-0.5$, $0$, $+0.5$, $+1$, and $+4$~\AA\ relative to the corresponding line center. Display ranges are selected to retain the morphological structure in both data sets and are not intended for quantitative comparison of absolute intensity between rows.}
\label{fig:data_context}
\end{figure}

\begin{figure}[!htbp]
\centering
\includegraphics[height=0.96\textheight,width=1.1\textwidth,keepaspectratio]{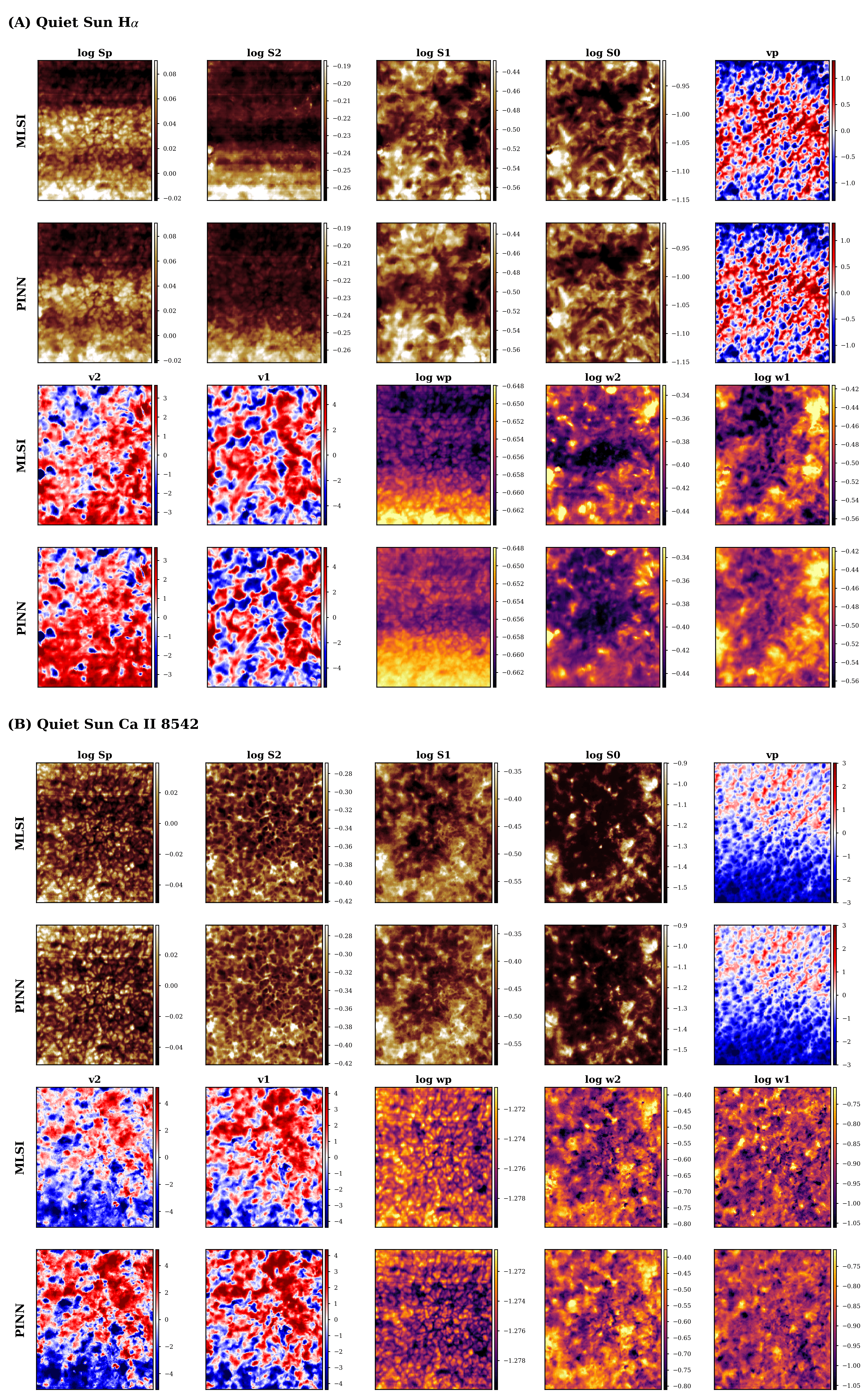}
\caption{Quiet-Sun parameter-map comparison for H$\alpha$ and Ca~II~8542. For each parameter, the conventional MLSI result and the MLSI-PINN prediction are shown using the same color scale.}
\label{fig:param_maps_qs}
\end{figure}

\subsection{FISS Observations and Spectral Morphology}
\label{subsec:results_observations}

Figure~\ref{fig:data_context} presents representative images from the quiet-Sun and active-region FISS observations. For both H$\alpha$ and Ca~II~8542, the panels sample the same seven wavelength offsets from $-4$ to $+4$~\AA. The line-continuum images mainly trace photospheric and lower-atmospheric structures, whereas the line-center images emphasize chromospheric morphology.

The quiet-Sun observations show relatively moderate contrast and spatially distributed fine structure. In comparison, the active-region images contain a prominent sunspot, stronger intensity gradients, and more structured line-core features. The active-region profiles therefore span a broader range of line depths, Doppler shifts, and asymmetries, providing a more demanding test of the inversion model. The wavelength-dependent changes in both regions also illustrate why the full spectral profile, rather than a single monochromatic image, is required for quantitative inversion.

Having established the observational morphology and spectral complexity of the two data sets, we next examine whether the MLSI-PINN model preserves the physical-parameter structure obtained from conventional MLSI.

\subsection{Parameter-Space Agreement with Conventional MLSI}
\label{subsec:results_parameters}

We first compare the spatial distributions of the inferred MLSI parameters. This comparison is important because a good spectral reconstruction alone does not guarantee that the inferred parameters reproduce the spatially coherent structures obtained by conventional MLSI.

Figure~\ref{fig:param_maps_qs} shows the quiet-Sun parameter maps for H$\alpha$ and Ca~II~8542. The MLSI-PINN predictions reproduce the large-scale morphology of the conventional MLSI maps for both lines. The source-function and Doppler-width parameters preserve the principal quiet-Sun contrast patterns, while the velocity-related quantities show somewhat larger local differences, consistent with their sensitivity to small line-center shifts and multilayer parameter degeneracy.

\begin{figure}[!p]
\centering
\includegraphics[height=0.96\textheight,width=1.1\textwidth,keepaspectratio]{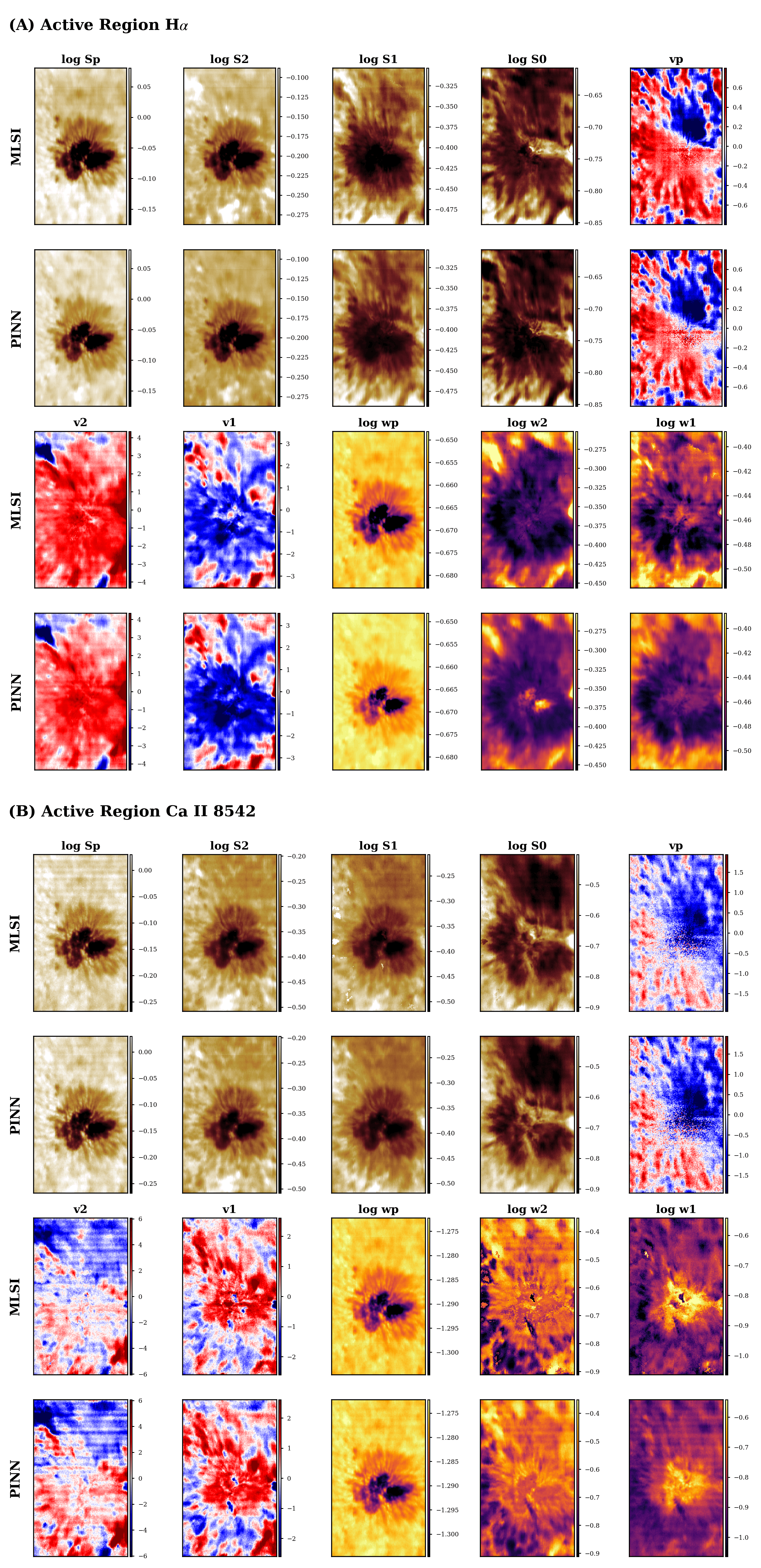}
\caption{Active-region parameter-map comparison for H$\alpha$ and Ca~II~8542. For each parameter, the conventional MLSI result is compared with the MLSI-PINN prediction using a shared color scale. The stronger spatial gradients in the active-region maps provide a more stringent test of whether the neural-network inversion preserves the parameter morphology of conventional MLSI.}
\label{fig:param_maps_ar}
\end{figure}

The active-region parameter maps are shown in Figure~\ref{fig:param_maps_ar}. Compared with the quiet-Sun case, the active-region maps contain sharper structures and stronger local variations, particularly in the line-core source functions and velocity-related parameters. Even under these more complex conditions, the MLSI-PINN results recover the dominant active-region morphology in both spectral lines. The agreement is strongest for parameters controlling the overall line depth and width, whereas the remaining differences are most apparent in quantities associated with local line asymmetries and small line-core shifts.

\begin{figure}[!htbp]
\centering
\includegraphics[width=\textwidth]{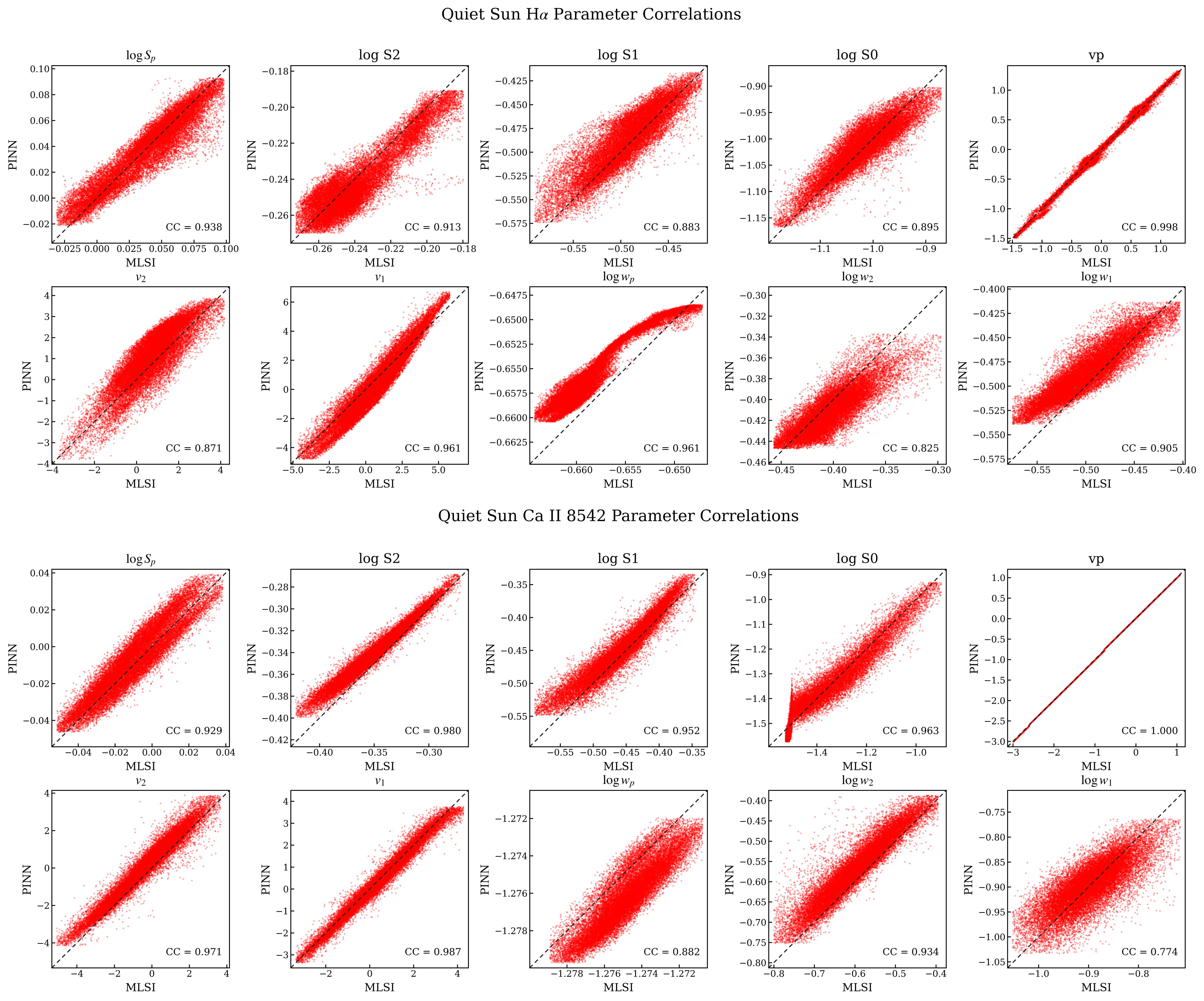}
\caption{Quiet-Sun parameter-space comparison between MLSI-PINN and conventional MLSI for one representative inference raster. The upper two rows show the H$\alpha$ parameters, and the lower two rows show the Ca~II~8542 parameters. The horizontal and vertical axes give the conventional MLSI and MLSI-PINN values, respectively. The dashed diagonal marks the one-to-one relation, and each panel reports the pixel-by-pixel Pearson correlation coefficient. The high agreement of $v_p$ primarily reflects the shared line-center proxy treatment.}
\label{fig:density_qs}
\end{figure}

\begin{figure}[!htbp]
\centering
\includegraphics[width=\textwidth]{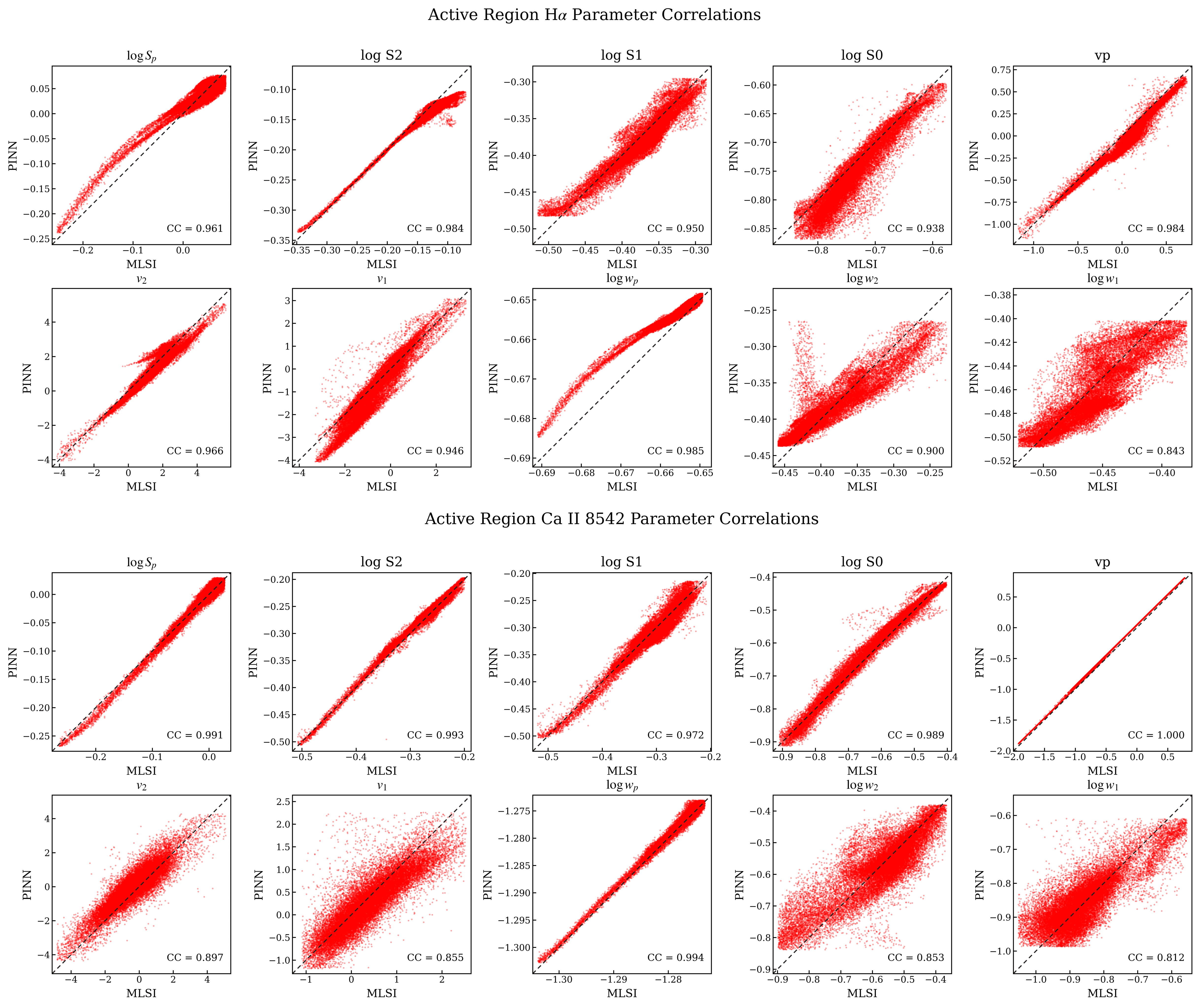}
\caption{Active-region parameter-space comparison between MLSI-PINN and conventional MLSI for one representative inference raster. The upper two rows show the H$\alpha$ parameters, and the lower two rows show the Ca~II~8542 parameters. The dashed diagonal indicates the one-to-one relation. The broader distributions relative to the quiet-Sun case reflect the larger range of line depths, Doppler shifts, and profile asymmetries in the active region.}
\label{fig:density_ar}
\end{figure}

To quantify the pixel-by-pixel agreement, Figures~\ref{fig:density_qs} and~\ref{fig:density_ar} compare the MLSI-PINN predictions with conventional MLSI parameters for representative quiet-Sun and active-region rasters. In each panel, the horizontal axis gives the direct MLSI value, the vertical axis gives the MLSI-PINN prediction, and the dashed diagonal indicates the one-to-one relation. The correlation coefficients reported in these figures are calculated from individual representative rasters, whereas Table~\ref{tab:param_metrics} reports statistics aggregated over all evaluated inference rasters. Small differences between the figure-level and sequence-level coefficients are therefore expected.

Figure~\ref{fig:density_qs} combines the quiet-Sun H$\alpha$ and Ca~II~8542 comparisons. The source-function parameters form compact, positively correlated distributions, with CC values of 0.883--0.938 for H$\alpha$ and 0.929--0.980 for Ca~II~8542. Lower agreement is found for H$\alpha$ $v_2$ and $\log w_2$, with CC values of 0.871 and 0.825, respectively, and for Ca~II $\log w_1$, with a CC of 0.774. The chromospheric velocities and selected Doppler widths exhibit broader distributions, consistent with their greater sensitivity to line-core shifts and degeneracy among multilayer parameters. Although the Doppler-width parameters generally have small absolute errors, their relatively narrow intrinsic ranges make the correlation and $R^2$ values more sensitive to small systematic offsets. The high agreement of $v_p$, with CC values of 0.998 for H$\alpha$ and 1.000 for Ca~II, primarily reflects the shared line-center proxy treatment rather than an independently learned neural-network prediction.

Figure~\ref{fig:density_ar} presents the corresponding active-region comparisons. The distributions span broader parameter ranges than in the quiet Sun, reflecting the stronger line-profile variability and spatial gradients of the active region. The source-function parameters retain high CC values of 0.938--0.984 for H$\alpha$ and 0.972--0.993 for Ca~II~8542. Greater scatter occurs in Ca~II $v_1$, $\log w_2$, and $\log w_1$, whose CC values are 0.855, 0.853, and 0.812, respectively, while H$\alpha$ $\log w_1$ has a CC of 0.843. This behavior is consistent with the reduced uniqueness of these chromospheric parameters within the multilayer forward model. Nevertheless, clear correlations remain for most parameters, particularly the source functions and photospheric Doppler widths. Overall, the active-region results indicate that the network preserves the main MLSI parameter-space structure even when the observed profiles contain stronger line asymmetries and a broader range of Doppler shifts. 

Together, Figures~\ref{fig:density_qs} and~\ref{fig:density_ar} show that MLSI-PINN reproduces the parameter-space structure of conventional MLSI for both spectral lines and solar targets. The CC values range from 0.774 to 1.000 for the representative quiet-Sun raster and from 0.812 to 1.000 for the active-region raster. The comparatively lower agreement is concentrated in chromospheric velocities and selected Doppler widths, which are less uniquely constrained by the emergent spectral profiles.

Table~\ref{tab:param_metrics} summarizes the parameter statistics and quantitative agreement with the conventional MLSI solution. The means and standard deviations are computed from the MLSI-PINN parameter maps, while the mean absolute error (MAE), Pearson correlation coefficient (CC), and coefficient of determination ($R^2$ score) are computed relative to the direct MLSI parameters. The table includes all evaluated inference rasters, whereas the CC values in Figures~\ref{fig:density_qs} and~\ref{fig:density_ar} correspond to single representative rasters.

The source-function parameters are recovered most consistently across both spectral lines and both solar targets. For the quiet-Sun case, the CC values of $\log S_p$, $\log S_2$, $\log S_1$, and $\log S_0$ are generally close to or above 0.9 for both H$\alpha$ and Ca~II~8542, and the corresponding MAEs remain small compared with the parameter dispersions. The same trend is observed in the active-region case, where the source-function parameters remain strongly correlated with the direct MLSI solution despite the broader range of line depths and local spectral asymmetries. This indicates that the spectral reconstruction loss and limited parameter supervision preserve the source-function structure inferred by conventional MLSI.

The photospheric velocity $v_p$ also shows very high agreement, but this parameter should be interpreted separately. In the MLSI-PINN output, $v_p$ is estimated from the observed line-center proxy, whereas in conventional MLSI it is obtained as part of the full-profile fitting. Both quantities are controlled primarily by the photospheric line-center shift and therefore remain highly consistent. However, because the two estimates are not produced by exactly the same calculation, the CC is not necessarily equal to unity. Small differences can arise from line-center measurement uncertainty, residual wavelength-calibration errors, finite spectral sampling, and local profile asymmetries.

The chromospheric velocity parameters exhibit larger scatter than the source-function parameters. This behavior is expected because $v_1$ and $v_2$ mainly affect line-core shifts and asymmetries, which can be partially compensated by changes in Doppler width, opacity, and source-function gradients in a multilayer model. Even so, the velocity correlations remain positive in all cases, showing that the MLSI-PINN captures the dominant velocity structure. The active-region H$\alpha$ case gives particularly strong agreement for $v_2$, while the Ca~II~8542 velocity parameters show larger errors, consistent with the broader distributions in Figure~\ref{fig:density_ar}.

The Doppler-width parameters show a different behavior. Their MAEs are generally small in absolute value, particularly for $\log w_p$, while some of their $R^2$ values are more modest. This does not necessarily imply a large physical discrepancy. Instead, it reflects the narrow intrinsic dynamic range of several width parameters: when the reference quantity varies only weakly across the field of view, even a small residual offset can noticeably reduce $R^2$. The width parameters should therefore be assessed using the MAE and CC together with $R^2$, rather than from the coefficient of determination alone.

\begin{table}[!htbp]
\centering
\scriptsize
\setlength{\tabcolsep}{4.0pt}
\renewcommand{\arraystretch}{1.08}

\vspace{0.2em}

\caption{Means and standard deviations of the MLSI-PINN physical parameters, with MAE, CC, and $R^2$ score relative to direct MLSI calculations.}
\label{tab:param_metrics}

\vspace{0.4em}
\hspace*{-0.12\textwidth}%
\resizebox{0.98\textwidth}{!}{%
\begin{tabular}{c c c c c c c c c c}
\hline
Region &
Physical Parameter &
\multicolumn{2}{c}{Mean $\pm$ Standard Deviation} &
\multicolumn{2}{c}{MAE} &
\multicolumn{2}{c}{CC} &
\multicolumn{2}{c}{$R^2$ Score} \\
\cline{3-4}\cline{5-6}\cline{7-8}\cline{9-10}
 & & H$\alpha$ & Ca~II & H$\alpha$ & Ca~II & H$\alpha$ & Ca~II & H$\alpha$ & Ca~II \\
\hline
\raisebox{-12.2ex}[0pt][0pt]{QS}
 & $\log S_p$ [$I_0$] & $0.037\pm0.028$ & $-0.006\pm0.020$ & 0.008 & 0.006 & 0.943 & 0.937 & 0.887 & 0.857 \\
 & $\log S_2$ [$I_0$] & $-0.240\pm0.020$ & $-0.340\pm0.027$ & 0.008 & 0.010 & 0.919 & 0.983 & 0.816 & 0.858 \\
 & $\log S_1$ [$I_0$] & $-0.488\pm0.035$ & $-0.446\pm0.049$ & 0.014 & 0.014 & 0.897 & 0.956 & 0.769 & 0.874 \\
 & $\log S_0$ [$I_0$] & $-1.020\pm0.055$ & $-1.374\pm0.154$ & 0.021 & 0.033 & 0.912 & 0.966 & 0.826 & 0.933 \\
 & $v_p$ [km s$^{-1}$] & $-0.011\pm0.626$ & $-0.747\pm0.957$ & 0.028 & 0.008 & 0.998 & 1.000 & 0.996 & 1.000 \\
 & $v_2$ [km s$^{-1}$] & $0.970\pm1.621$ & $0.018\pm2.612$ & 0.628 & 0.845 & 0.907 & 0.972 & 0.756 & 0.664 \\
 & $v_1$ [km s$^{-1}$] & $0.386\pm2.549$ & $0.528\pm2.032$ & 0.629 & 0.390 & 0.964 & 0.986 & 0.884 & 0.911 \\
 & $\log w_p$ [\AA] & $-0.655\pm0.003$ & $-1.276\pm0.002$ & 0.003 & 0.001 & 0.961 & 0.894 & 0.550 & 0.233 \\
 & $\log w_2$ [\AA] & $-0.409\pm0.024$ & $-0.566\pm0.083$ & 0.014 & 0.034 & 0.855 & 0.938 & 0.644 & 0.750 \\
 & $\log w_1$ [\AA] & $-0.483\pm0.030$ & $-0.895\pm0.053$ & 0.018 & 0.030 & 0.913 & 0.799 & 0.676 & 0.638 \\
\hline
\raisebox{-12.2ex}[0pt][0pt]{AR}
 & $\log S_p$ [$I_0$] & $0.034\pm0.056$ & $-0.024\pm0.059$ & 0.015 & 0.006 & 0.969 & 0.992 & 0.922 & 0.981 \\
 & $\log S_2$ [$I_0$] & $-0.147\pm0.043$ & $-0.281\pm0.057$ & 0.009 & 0.006 & 0.985 & 0.995 & 0.933 & 0.984 \\
 & $\log S_1$ [$I_0$] & $-0.382\pm0.043$ & $-0.315\pm0.057$ & 0.013 & 0.010 & 0.951 & 0.971 & 0.888 & 0.943 \\
 & $\log S_0$ [$I_0$] & $-0.753\pm0.065$ & $-0.639\pm0.129$ & 0.023 & 0.015 & 0.938 & 0.989 & 0.727 & 0.975 \\
 & $v_p$ [km s$^{-1}$] & $0.000\pm0.368$ & $-0.250\pm0.757$ & 0.080 & 0.113 & 0.975 & 1.000 & 0.913 & 0.936 \\
 & $v_2$ [km s$^{-1}$] & $1.258\pm1.462$ & $0.025\pm2.116$ & 0.317 & 0.723 & 0.977 & 0.922 & 0.932 & 0.659 \\
 & $v_1$ [km s$^{-1}$] & $-0.889\pm1.458$ & $0.404\pm0.877$ & 0.417 & 0.321 & 0.946 & 0.846 & 0.805 & 0.643 \\
 & $\log w_p$ [\AA] & $-0.653\pm0.006$ & $-1.278\pm0.006$ & 0.002 & 0.001 & 0.986 & 0.994 & 0.845 & 0.984 \\
 & $\log w_2$ [\AA] & $-0.382\pm0.042$ & $-0.580\pm0.104$ & 0.017 & 0.048 & 0.906 & 0.857 & 0.799 & 0.733 \\
 & $\log w_1$ [\AA] & $-0.461\pm0.030$ & $-0.847\pm0.085$ & 0.014 & 0.046 & 0.851 & 0.809 & 0.705 & 0.651 \\
\hline
\end{tabular}%
}
\end{table}

\FloatBarrier
\subsection{Spectral Reconstruction and Temporal Evolution}
\label{subsec:results_spectra_time}

The parameter-space comparison demonstrates that the network preserves the main MLSI parameter structure. We now examine whether these inferred parameters also reproduce the observed spectral profiles and their temporal evolution.

\begin{figure}[!htbp]
\centering
\includegraphics[width=\textwidth]{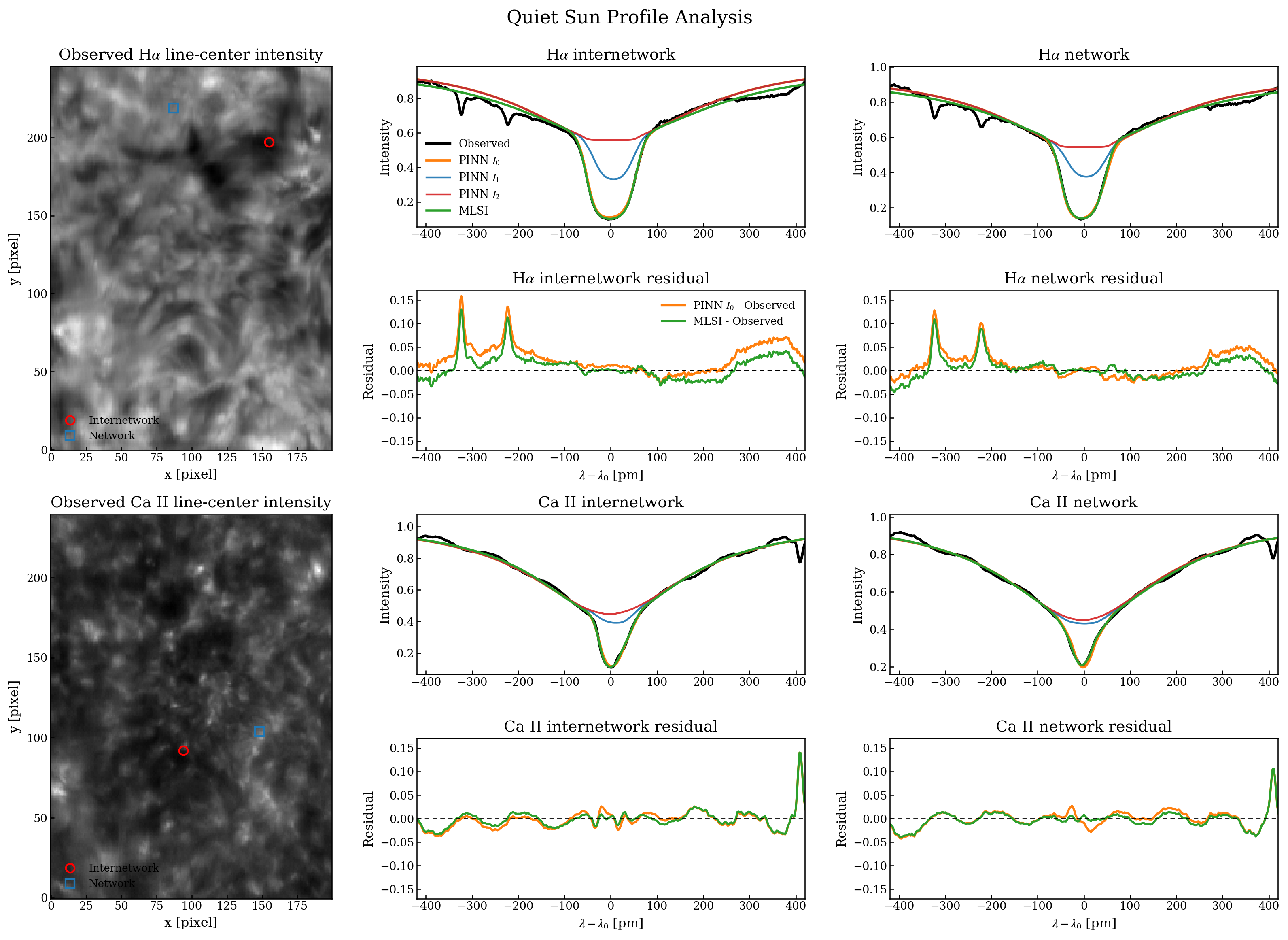}
\caption{Quiet-Sun spectral-profile comparison for representative internetwork and network pixels. The left panels show the observed H$\alpha$ and Ca~II~8542 line-center intensity maps with the selected locations marked. The right panels compare the observed profiles with the MLSI-PINN emergent profile $I_0$, the intermediate profiles $I_1$ and $I_2$, and the conventional MLSI reconstruction. The residual panels show $I_0-\mathrm{Observed}$ and $\mathrm{MLSI}-\mathrm{Observed}$. The wavelength offset is defined relative to the corresponding line center of each transition.}
\label{fig:spectral_qs_profiles}
\end{figure}

\begin{figure}[!htbp]
\centering
\includegraphics[width=\textwidth]{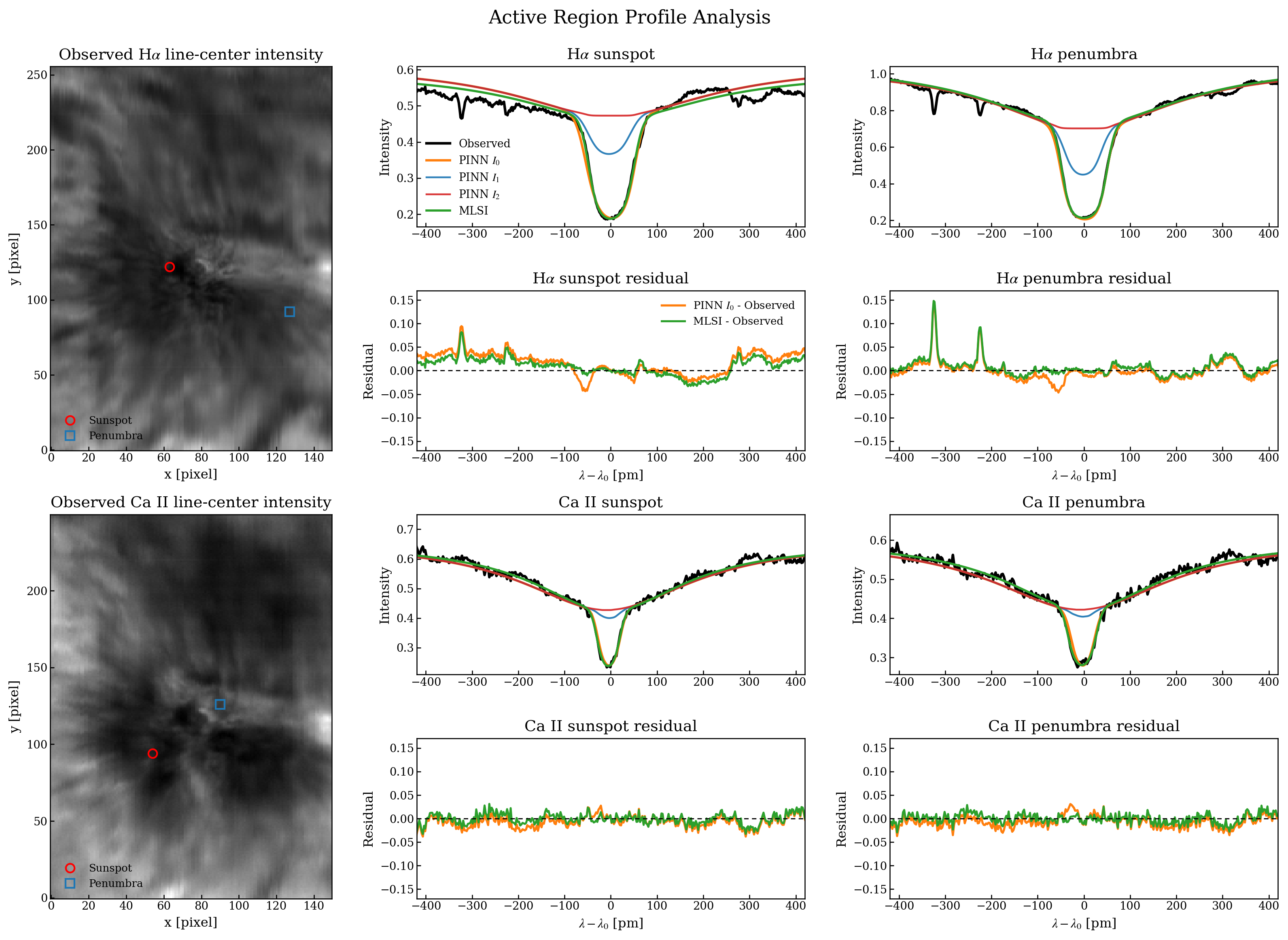}
\caption{Active-region spectral-profile comparison for representative sunspot umbral and penumbral pixels. The left panels show the observed H$\alpha$ and Ca~II~8542 line-center intensity maps with the selected locations marked. The right panels compare the observed profiles with the MLSI-PINN emergent profile $I_0$, the intermediate profiles $I_1$ and $I_2$, and the conventional MLSI reconstruction. The residual panels show $I_0-\mathrm{Observed}$ and $\mathrm{MLSI}-\mathrm{Observed}$. The wavelength offset is defined relative to the corresponding line center of each transition.}
\label{fig:spectral_ar_profiles}
\end{figure}

We first examine the reconstructed spectral profiles at representative quiet-Sun and active-region locations in Figures~\ref{fig:spectral_qs_profiles} and~\ref{fig:spectral_ar_profiles}, respectively. For each spectral line, the left panel shows the observed line-center intensity map from which representative pixels were selected. The quiet-Sun pixels represent internetwork and network regions, whereas the active-region pixels represent sunspot umbral and penumbral regions. Each plotted line profile is extracted from the individual pixel marked in the corresponding map, without spatial averaging. The right panels compare the observed line profile with the final MLSI-PINN emergent profile $I_0$, the intermediate layer profiles $I_1$ and $I_2$, and the conventional MLSI reconstruction. The intermediate profiles are not fitted to the observations independently; they illustrate how the embedded multilayer forward model constructs the final emergent line profile from the predicted atmospheric parameters. The wavelength axis is expressed as $\lambda-\lambda_0$ separately for each spectral line.

Figure~\ref{fig:spectral_qs_profiles} shows the quiet-Sun profile comparison for representative internetwork and network locations. For both H$\alpha$ and Ca~II~8542, the final MLSI-PINN line profile reproduces the main absorption profile and closely follows the conventional MLSI reconstruction, particularly around the line core. To avoid contamination from the outer continuum regions, the residual statistics were calculated only over the central wavelength interval from $-200$ to $+200$~pm. Defining the residual as the reconstructed intensity minus the observed intensity, the H$\alpha$ mean residual averaged over the two selected pixels is $2.31\times10^{-3}$ for MLSI-PINN and $-5.66\times10^{-4}$ for conventional MLSI. The corresponding mean absolute residuals are $1.24\times10^{-2}$ and $1.05\times10^{-2}$ in normalized-intensity units. For Ca~II~8542, the mean residuals are $2.10\times10^{-3}$ for MLSI-PINN and $9.28\times10^{-5}$ for conventional MLSI, while the mean absolute residuals are $9.95\times10^{-3}$ and $7.59\times10^{-3}$, respectively. The positive MLSI-PINN mean residuals indicate a small average upward intensity bias in the quiet-Sun reconstructions, although the offset is not uniform across every wavelength or selected pixel. The slightly lower residuals of conventional MLSI are expected because its parameters are optimized independently for each observed line profile. Localized residual features appearing in both reconstructions are more likely associated with observational spectral structure and the finite flexibility of the compact MLSI forward model. 

Figure~\ref{fig:spectral_ar_profiles} shows the corresponding active-region comparison for representative sunspot umbral and penumbral locations. Compared with the quiet-Sun profiles, the active-region spectra exhibit stronger local variations in their line cores and wings. All residual panels in Figures~\ref{fig:spectral_qs_profiles} and~\ref{fig:spectral_ar_profiles} use the same vertical range of $-0.17$ to $0.17$, allowing their amplitudes to be compared directly. Using the same $-200$ to $+200$~pm interval, the H$\alpha$ mean residuals are $-5.02\times10^{-3}$ for MLSI-PINN and $-5.60\times10^{-4}$ for conventional MLSI, with mean absolute residuals of $1.27\times10^{-2}$ and $8.67\times10^{-3}$, respectively. For Ca~II~8542, the corresponding mean residuals are $-5.18\times10^{-3}$ and $-1.30\times10^{-3}$, while the mean absolute residuals are $1.08\times10^{-2}$ and $7.03\times10^{-3}$. In contrast to the small positive bias found in the quiet-Sun examples, the selected active-region profiles show a modest negative MLSI-PINN intensity bias. Nevertheless, the mean absolute residuals remain of the same order for the two data sets, and MLSI-PINN continues to recover the dominant absorption morphology in both spectral lines. These results indicate that the network retains the physically interpretable MLSI representation with a modest additional reconstruction error relative to independently optimized conventional MLSI.

We also examine the temporal evolution of the reconstructed spectra and physical parameters in Figures 9-12. For each data set, the time--wavelength diagrams compare the observed line profile, the conventional MLSI reconstruction, the MLSI-PINN reconstruction, and their residuals. The corresponding parameter time series show the upper-layer line-of-sight velocities $v_1$ inferred from H$\alpha$ and Ca~II~8542, together with two derived quantities, the temperature $T$ and nonthermal velocity $\xi$.

The temperature and nonthermal velocity are not independent MLSI-PINN output parameters. They are derived from the upper-chromospheric Doppler widths of the two lines, assuming that the H$\alpha$ and Ca~II~8542 upper-layer widths sample the same plasma and that the line broadening can be decomposed into thermal and nonthermal components:
\begin{equation}
\left(\frac{c w_j}{\lambda_j}\right)^2
=
\frac{2k_{\rm B}T}{m_j}+\xi^2,
\qquad j=\mathrm{H}\alpha,\ \mathrm{Ca~II}.
\end{equation}
Using $w_{\rm H}$ and $w_{\rm Ca}$ for the upper-layer Doppler widths $w_1$ of H$\alpha$ and Ca~II~8542, respectively, this gives
\begin{equation}
T =
8100
\left(\frac{w_{\rm H}}{0.025~{\rm nm}}\right)^2
\left[
1 - 0.59
\left(\frac{w_{\rm Ca}}{w_{\rm H}}\right)^2
\right]~{\rm K},
\end{equation}
and
\begin{equation}
\xi =
5.40
\left(\frac{w_{\rm Ca}}{0.015~{\rm nm}}\right)
\left[
1 - 0.042
\left(\frac{w_{\rm H}}{w_{\rm Ca}}\right)^2
\right]^{1/2}
~{\rm km~s^{-1}} .
\end{equation}

\begin{figure}[!htbp]
\centering
\includegraphics[width=\textwidth]{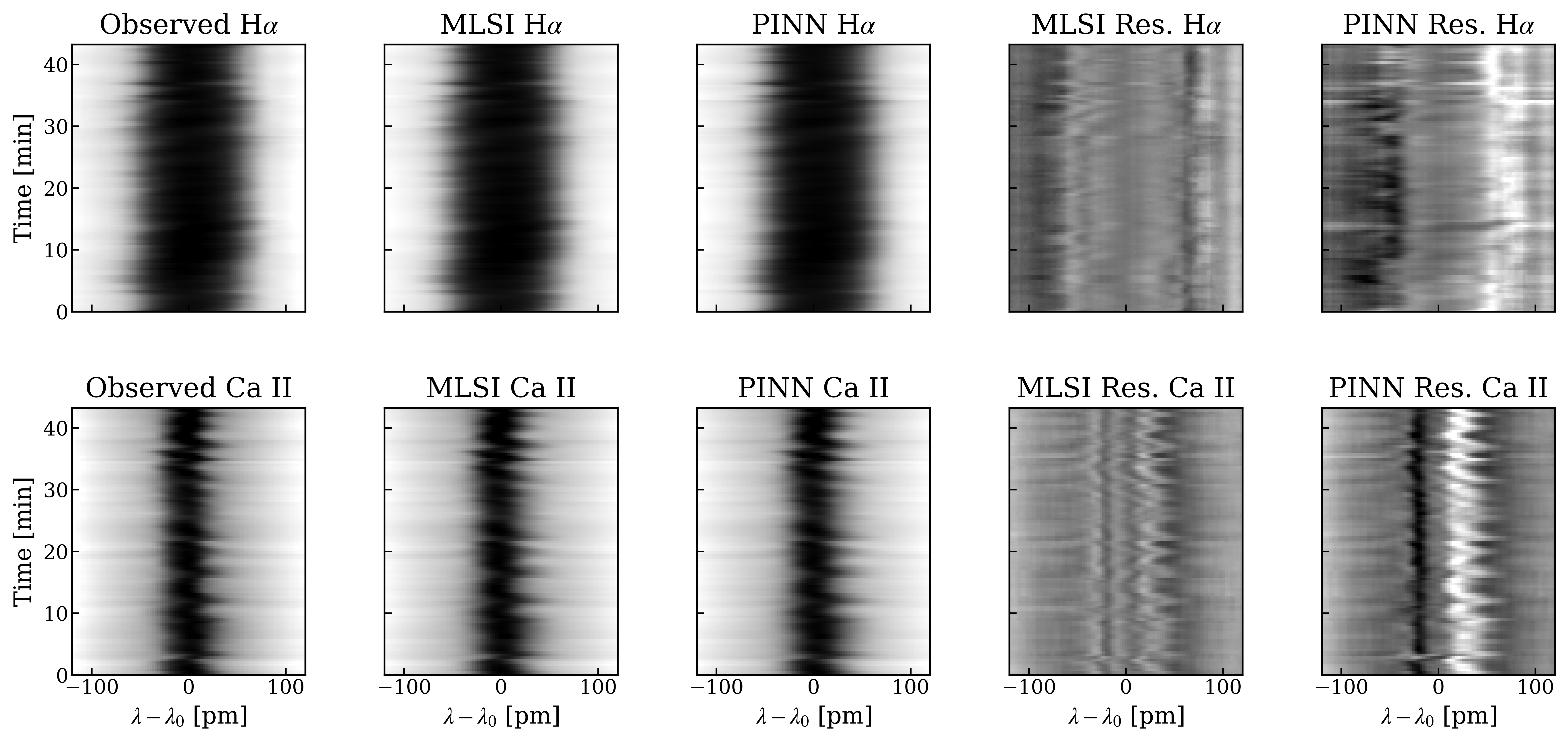}
\caption{Quiet-Sun time--wavelength spectral comparison for H$\alpha$ and Ca~II~8542. The columns show the observed spectra, the conventional MLSI reconstruction, the MLSI-PINN reconstruction, and the corresponding MLSI and MLSI-PINN residuals.}
\label{fig:time_qs_spec}
\end{figure}

Figure~\ref{fig:time_qs_spec} shows the quiet-Sun time--wavelength comparison. Both H$\alpha$ and Ca~II~8542 show that the MLSI-PINN reconstruction follows the observed temporal evolution of the line profile and remains close to the conventional MLSI reconstruction.

\begin{figure}[!htbp]
\centering
\includegraphics[width=0.55\textwidth]{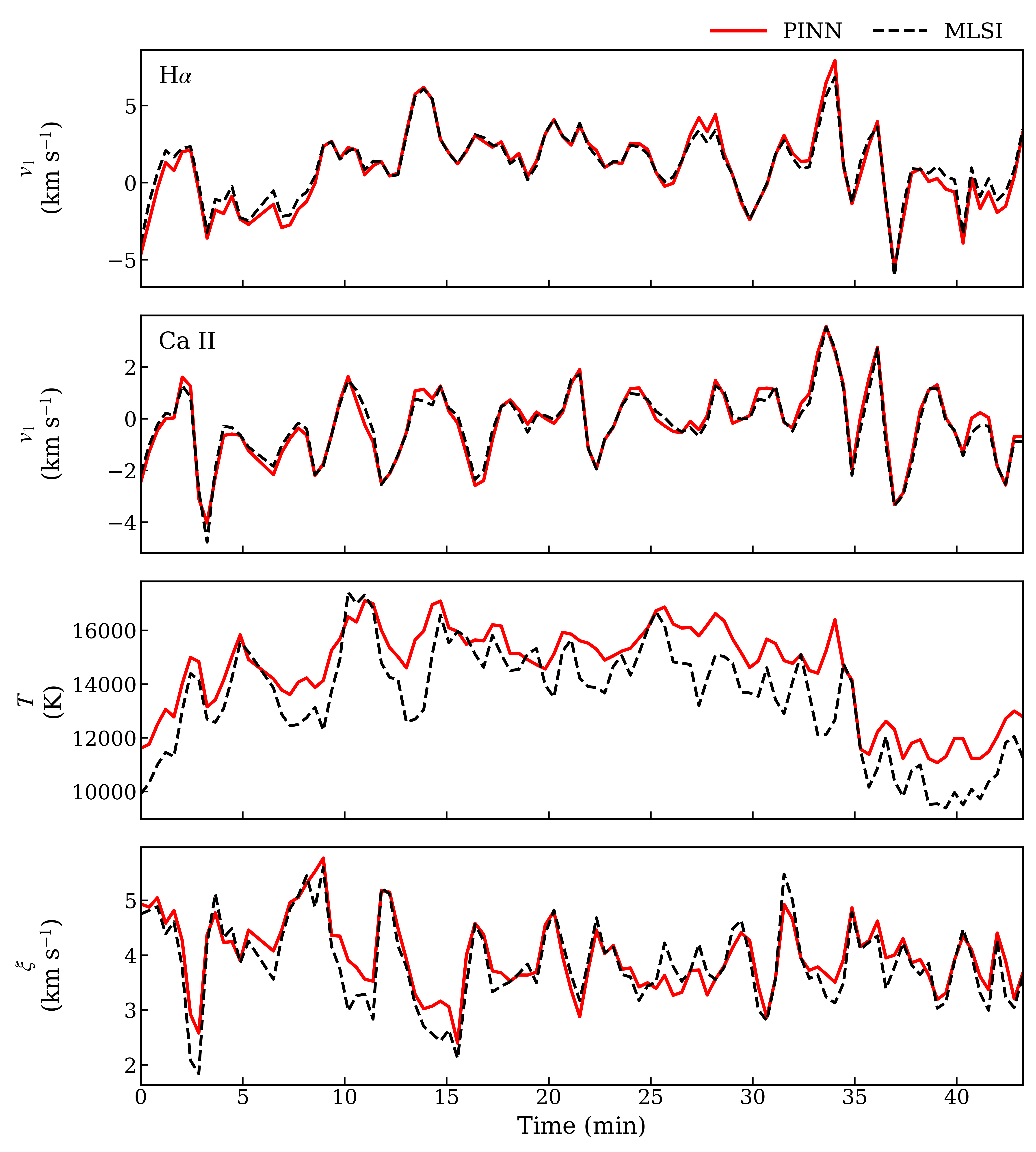}
\caption{Quiet-Sun temporal evolution of the upper-layer velocities $v_1$ from H$\alpha$ and Ca~II~8542, and the derived temperature $T$ and nonthermal velocity $\xi$.}
\label{fig:time_qs_param}
\end{figure}

Figure~\ref{fig:time_qs_param} shows the quiet-Sun temporal evolution of $v_1$, $T$, and $\xi$. The MLSI-PINN results closely track the MLSI reference for both line-of-sight velocities and for the derived thermal and nonthermal quantities, indicating that the network preserves the temporal behavior of the MLSI solution.

\begin{figure}[!htbp]
\centering
\includegraphics[width=\textwidth]{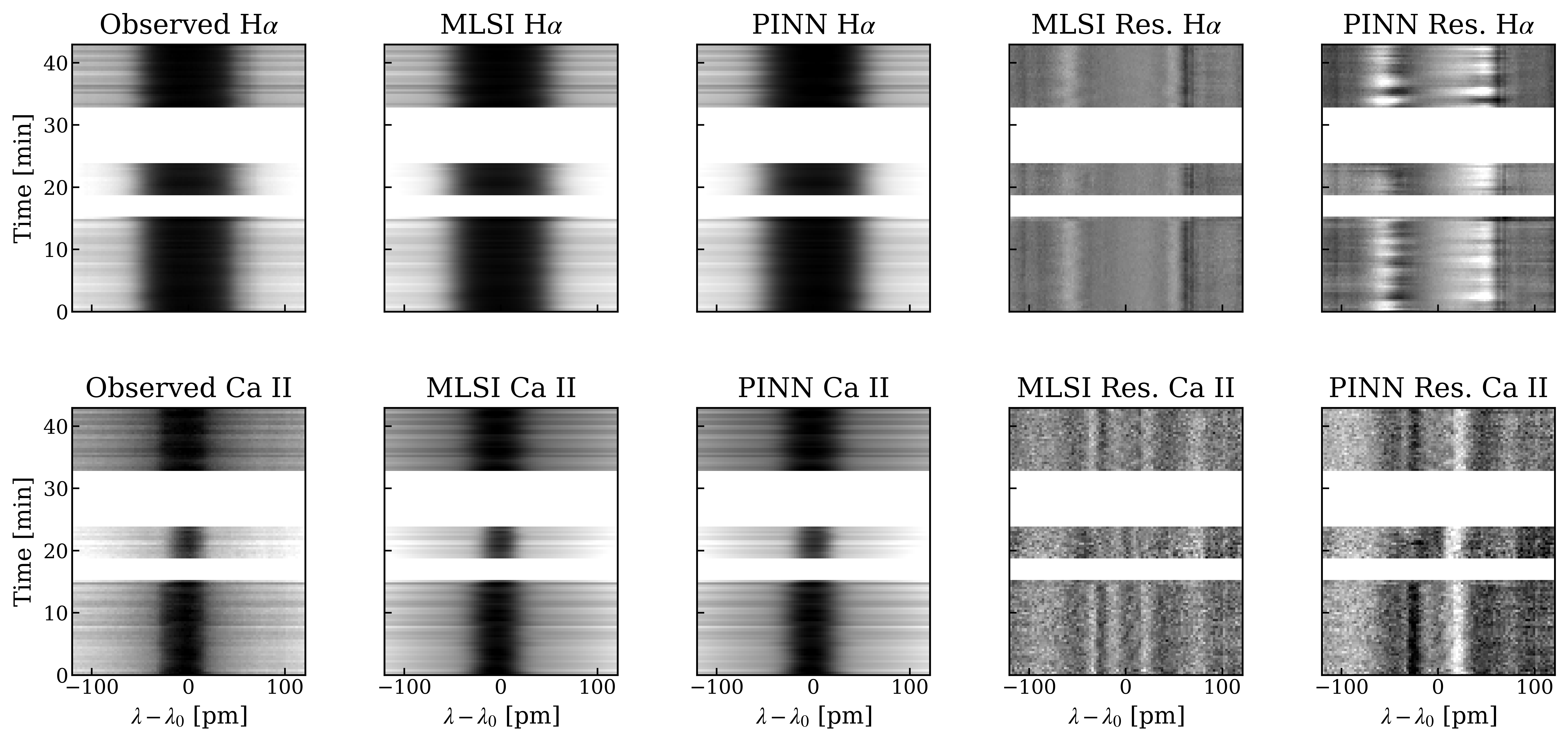}
\caption{Active-region time--wavelength spectral comparison for H$\alpha$ and Ca~II~8542. The columns show the observed spectra, the conventional MLSI reconstruction, the MLSI-PINN reconstruction, and the corresponding MLSI and MLSI-PINN residuals. White horizontal bands indicate missing observing intervals.}
\label{fig:time_ar_spec}
\end{figure}

\begin{figure}[!htbp]
\centering
\includegraphics[width=0.55\textwidth]{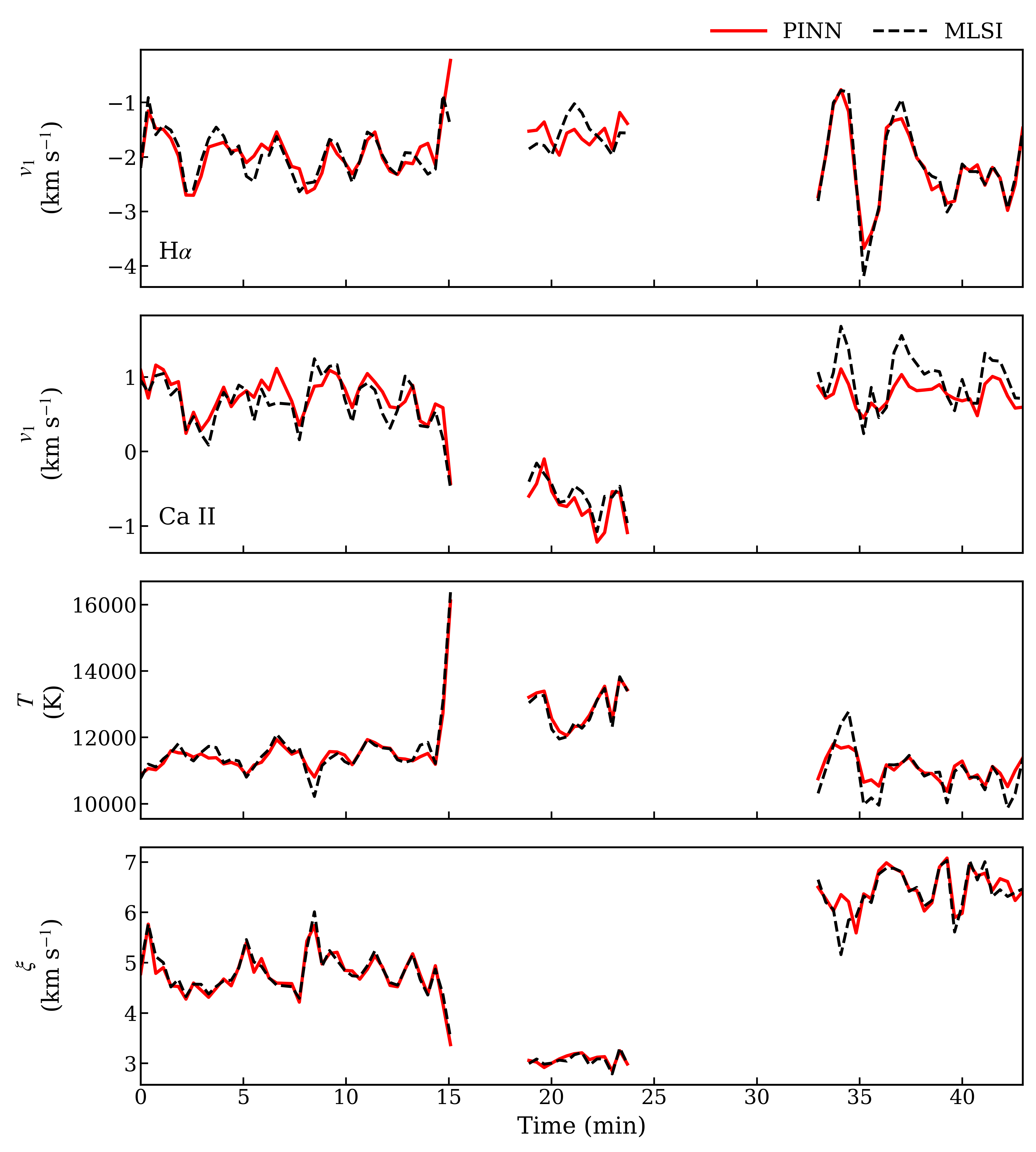}
\caption{Active-region temporal evolution of $v_1$, $T$, and $\xi$. The curves are broken across missing observing intervals, so no interpolation is implied across the gaps.}
\label{fig:time_ar_param}
\end{figure}

Figures~\ref{fig:time_ar_spec} and~\ref{fig:time_ar_param} show the corresponding active-region results. The active-region sequence contains observational gaps, which are retained in the time axis rather than interpolated. These missing intervals appear as white horizontal bands in the time--wavelength diagrams and as breaks in the parameter time series. Within the observed intervals, MLSI-PINN follows the main time-dependent line-profile changes and reproduces the MLSI temporal trends in the upper-layer velocities and derived quantities.

\begin{figure*}[!htbp]
\centering
\includegraphics[width=\textwidth]{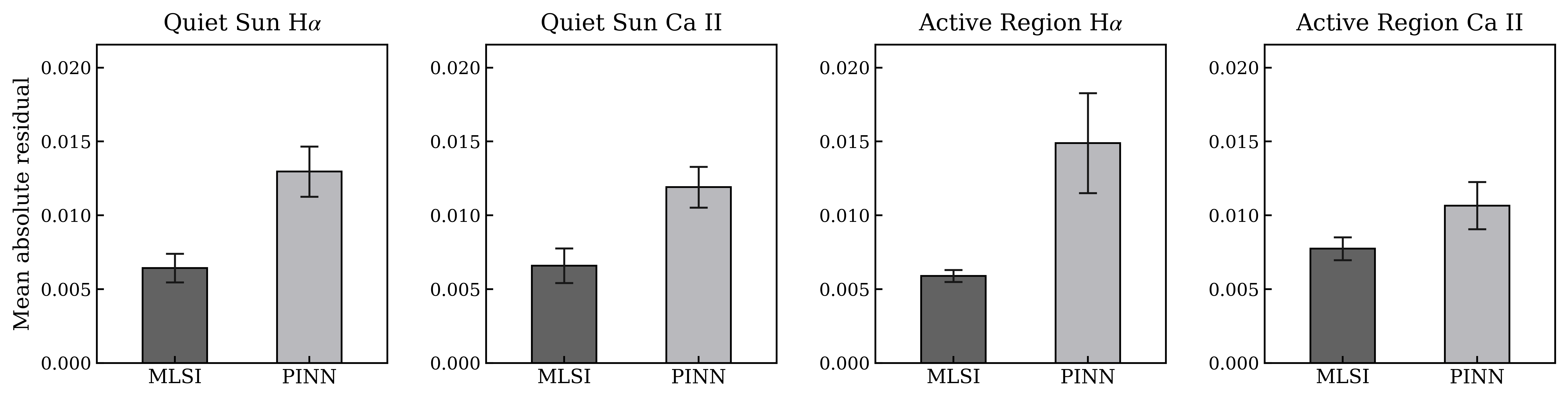}
\caption{Summary of the spectral reconstruction residuals in Figures~\ref{fig:time_qs_spec} and~\ref{fig:time_ar_spec}. The four panels show quiet-Sun H$\alpha$, quiet-Sun Ca~II~8542, active-region H$\alpha$, and active-region Ca~II~8542, respectively. For each time step, the mean absolute residual is calculated over the displayed wavelength interval of $\lvert\lambda-\lambda_0\rvert\leq120$~pm. The bar height gives the temporal mean, and the error bar indicates one standard deviation over the observed sequence.}
\label{fig:time_residual_summary}
\end{figure*}

To provide a more direct quantitative comparison of the reconstruction accuracy, Figure~\ref{fig:time_residual_summary} summarizes the residuals shown in Figures~\ref{fig:time_qs_spec} and~\ref{fig:time_ar_spec}. The conventional MLSI mean absolute residuals range from approximately $5.9\times10^{-3}$ to $7.7\times10^{-3}$, while the MLSI-PINN values range from approximately $1.1\times10^{-2}$ to $1.5\times10^{-2}$. The error bars indicate the temporal standard deviation of the wavelength-averaged absolute residual, with the largest variation occurring for active-region H$\alpha$. Although the MLSI-PINN residuals are systematically higher, both methods remain on the same order of magnitude for the two spectral lines and solar targets. As seen in Figures~\ref{fig:time_qs_spec} and~\ref{fig:time_ar_spec}, the stronger MLSI-PINN residuals remain concentrated near similar wavelength positions throughout the sequences, particularly along the steep line-core flanks. A small difference in the reconstructed line-center position or profile width can produce a relatively large pointwise intensity residual where $\left|\partial I/\partial\lambda\right|$ is large. Consequently, the observed and reconstructed profiles can remain close in their overall morphology even when a slight spectral displacement produces coherent residual bands. Because the residual is evaluated at fixed wavelength samples, this metric penalizes small spectral shifts more strongly than a comparison based primarily on the overall profile shape. This distinction is important when interpreting the residual magnitude because it separates a small wavelength-position offset from a substantial error in reproducing the evolution of the line profile. This sensitivity is most evident for active-region H$\alpha$, which exhibits both the largest mean MLSI-PINN residual and the greatest temporal variation among the four cases. The residual structure is therefore consistent with small systematic differences in the inferred velocity or width parameters rather than purely random temporal reconstruction errors. Overall, MLSI-PINN maintains reconstruction accuracy comparable in scale to conventional MLSI, although the direct nonlinear optimization provides a closer pointwise fit to the observed spectra.

\FloatBarrier
\subsection{Computational Performance}
\label{subsec:results_runtime}

The preceding comparisons show that the MLSI-PINN model retains the main parameter-space and spectral-reconstruction behavior of conventional MLSI. The practical value of the proposed framework also depends on its computational cost. We therefore record approximate wall-clock times for conventional MLSI and for the MLSI-PINN workflow. The conventional MLSI baseline is CPU-based and performs pixel-by-pixel nonlinear least-squares fitting independently for each raster. In contrast, MLSI-PINN requires a one-time two-stage training step on a reference raster, after which each additional raster is processed by neural-network inference followed by the analytic MLSI forward calculation. The reported MLSI-PINN inference time includes parameter prediction and spectral forward synthesis, but excludes file I/O.

The timing comparison should be interpreted as a practical wall-clock comparison rather than a hardware-independent algorithmic benchmark. Conventional MLSI is implemented as a CPU-based pixel-by-pixel nonlinear fitting workflow, whereas MLSI-PINN uses GPU-accelerated neural-network inference after a one-time training step. A fully hardware-matched comparison is therefore not straightforward. Nevertheless, this comparison reflects the practical deployment scenario: direct MLSI fitting requires several minutes per raster, while the trained MLSI-PINN model processes each additional raster in several seconds. For observing sequences containing tens to hundreds of rasters, the one-time training cost is amortized over many frames, making the proposed framework substantially more efficient for same-day FISS time-series analysis.

\begin{table}[!htbp]
\centering
\caption{Approximate wall-clock computational cost of conventional MLSI and MLSI-PINN.}
\label{tab:runtime}
\begin{tabular}{lcccc}
\hline
Method & Hardware & Training time & Inference time per raster & Notes \\
\hline
Conventional MLSI & CPU-based & -- & 3--5 min & Pixel-by-pixel nonlinear fitting \\
MLSI-PINN & RTX 5090 GPU & $\sim$15 min & 5--15 s & Two-stage training, then forward inference \\
\hline
\end{tabular}
\end{table}

Overall, the parameter maps, density diagrams, spectral reconstructions, temporal comparisons, and quantitative metrics give a consistent picture. The MLSI-PINN model reproduces the main spatial structures and parameter scales of the conventional MLSI inversion for both quiet-Sun and active-region observations. The agreement is strongest for the source-function parameters, while the $v_p$ agreement mainly reflects the consistency between the line-center proxy and the conventional MLSI photospheric-velocity estimate. The chromospheric velocity and Doppler-width parameters show larger deviations because they are more sensitive to line asymmetries, narrow dynamic ranges, and multilayer parameter degeneracy. Despite these limitations, the reconstructed spectra remain close to the observations and show residuals comparable to conventional MLSI forward reconstruction, indicating that the model preserves the main physical constraints of the MLSI formulation while avoiding repeated pixel-by-pixel nonlinear fitting.

\subsection{Spectral Reconstruction Using the Institut f\"ur Astrophysik G\"ottingen (IAG) Solar Flux Atlas}
\label{subsec:IAG}
As an additional test of spectral reconstruction beyond the FISS observations, we applied the stage-1 MLSI-PINN procedure to the H$\alpha$ and Ca~II~8542 profiles extracted from the IAG solar flux atlas \citep{Reiners2016}. The IAG profiles were resampled onto the corresponding FISS wavelength grids before training. As no reference MLSI parameters are available for these spectra, only the spectral reconstruction loss was used, without the stage-2 parameter-space supervision.

\begin{figure}[!htbp]
\centering
\includegraphics[width=\textwidth]{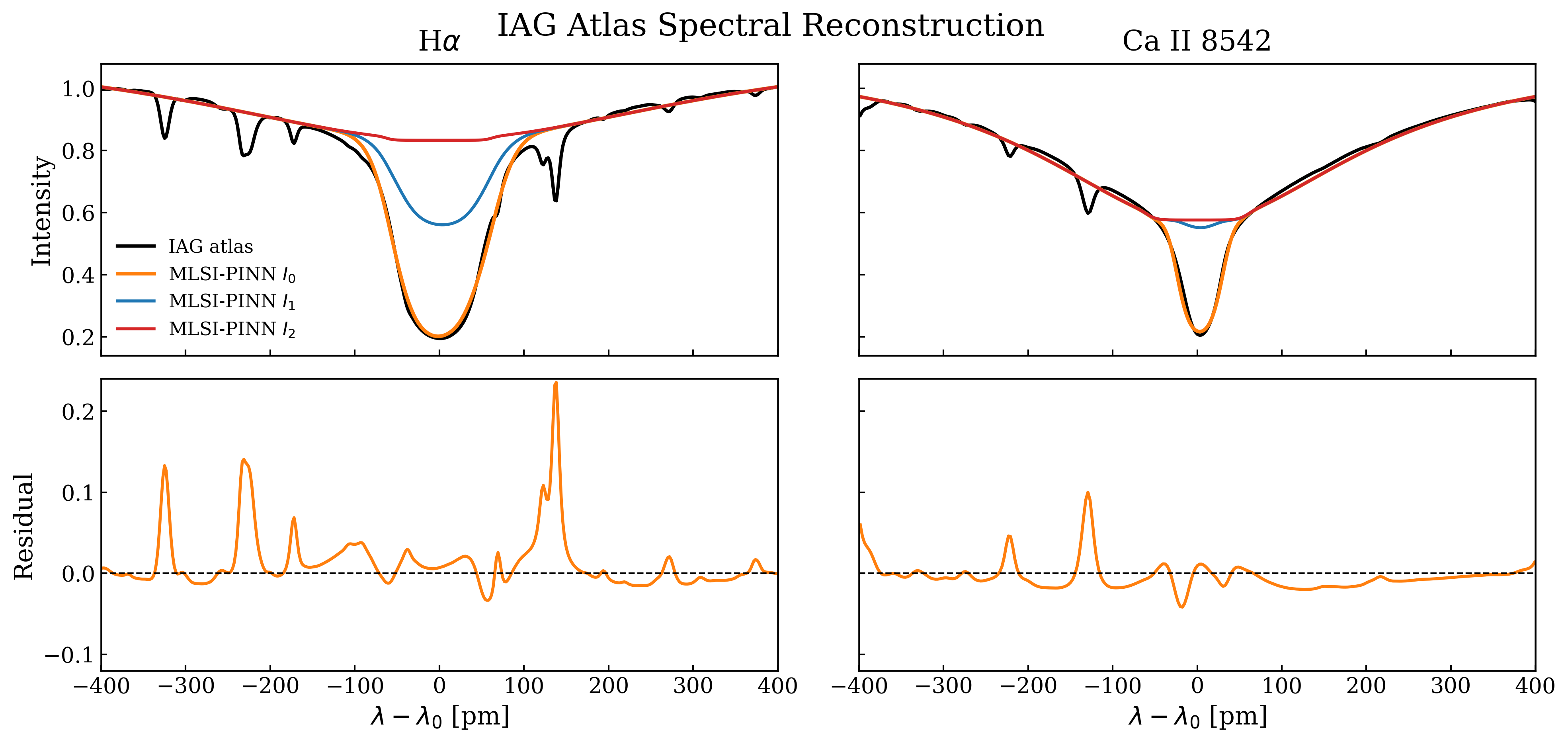}
\caption{Stage-1 MLSI-PINN reconstruction of the H$\alpha$ and Ca~II~8542 profiles extracted from the IAG solar flux atlas.}
\label{fig:iag_profiles}
\end{figure}

Figure~\ref{fig:iag_profiles} shows the observed IAG profiles together with the reconstructed emergent spectrum $I_0$ and the intermediate profiles $I_1$ and $I_2$. For both lines, $I_0$ reproduces the broad absorption profile and the line core. Within the clean wavelength samples over $\pm4$~\AA, the mean residuals, defined as $\langle I_0-I_{\mathrm{obs}}\rangle$, are $4.99\times10^{-3}$ for H$\alpha$ and $-6.53\times10^{-3}$ for Ca~II~8542. The prominent localized residual peaks, particularly in H$\alpha$, occur where narrow absorption features are not represented by the compact MLSI forward model. These features occupy only a small fraction of the wavelength interval and therefore do not characterize the overall reconstruction agreement.

This test demonstrates that the MLSI-PINN framework can be adapted to independently obtained, high-resolution solar spectra while retaining the same analytic multilayer construction. However, because the IAG atlas provides disk-integrated solar flux rather than reference atmospheric parameters, this comparison evaluates spectral representability rather than the accuracy of the inferred MLSI parameters.

\section{Summary and Discussion}
\label{sec:discussion}

In this paper, we have developed a two-stage physics-informed neural-network framework for accelerating multilayer spectral inversion of H$\alpha$ and Ca~II~8542 spectra. The method keeps the analytic MLSI forward model inside the training loop, so the network is not trained as a purely empirical mapping from spectra to labels. Instead, the predicted parameters are required to reproduce the observed spectra through the same multilayer radiative-transfer formulation used by MLSI. Applied to quiet-Sun and active-region FISS observations, the method reproduces the main parameter structures of conventional MLSI and generates reconstructed spectra with residuals comparable to direct MLSI forward synthesis.

The two-stage design is important for reducing parameter degeneracy. In stage 1, the model is trained only through the spectral reconstruction loss, which allows it to learn a fast inverse mapping directly from the observed profiles. However, the emergent line profile does not uniquely determine every MLSI parameter. Different combinations of source functions, velocities, optical depths, and Doppler widths can produce similar line profiles. Stage 2 therefore introduces limited parameter-space supervision from conventional MLSI results for a selected subset of parameters while retaining the spectral reconstruction loss. This provides additional guidance for parameters that are less strongly constrained by the line profile alone, without requiring a large precomputed training set.

The parameter-dependent performance follows the expected sensitivity of the MLSI forward model. Source-function parameters are more robust because they control the line depth and the broad profile shape. Chromospheric velocities and some Doppler-width parameters are more difficult because they affect subtler profile asymmetries and can be partially degenerate with other layer parameters. The photospheric velocity $v_p$ should be interpreted separately, since it is inserted from a line-center proxy rather than independently learned by the network. Its high agreement with conventional MLSI therefore reflects the consistency of the proxy estimate with the fitted MLSI photospheric velocity.

The present validation is focused on same-day or same-sequence application. This is a practical scope for FISS time-series analysis, because the wavelength sampling, calibration, and intensity normalization are relatively consistent within a sequence. Since the two-stage training cost is modest compared with repeated pixel-by-pixel MLSI fitting over many images, retraining or fine-tuning for a new observing day remains practical. The method should therefore be viewed as an efficient sequence-level acceleration strategy rather than a universal inversion model that can be transferred without adjustment across instruments, observing conditions, or substantially different calibration states.

However, there are still several limitations. The parameter supervision in stage 2 inherits the assumptions and possible biases of the conventional MLSI reference inversion. In addition, MLSI itself is a simplified analytic radiative-transfer model and cannot replace full NLTE inversions when detailed atmospheric stratification, magnetic-field information, or more realistic radiative transfer is required. Future work should test the framework on a broader set of observing days and solar targets, quantify uncertainty for parameters affected by degeneracy, and examine its transferability across different observations. As the core inversion pipeline ``mapping an observed spectrum to MLSI parameters and constraining them through the embedded analytic forward model" remains consistent across data sets, the framework is not inherently limited to the FISS sequences examined here. This consistency suggests that the method could, in principle, be extended to high-volume observations from DKIST. Investigating this possibility is a promising direction for future work, since rapid raster-scale inference could make MLSI-type analysis of large DKIST rasters and time sequences more practical. Despite these limitations, the present results show that MLSI-PINN can substantially reduce the practical cost of MLSI analysis while preserving the physical interpretability of the multilayer inversion parameters.

\begin{acknowledgments}
We gratefully acknowledge the use of data from the Goode Solar Telescope (GST) at Big Bear Solar Observatory (BBSO). BBSO operation is supported by NSF grant AGS-2309939 and the New Jersey Institute of Technology. GST operation is partly supported by the Korea Astronomy and Space Science Institute and Seoul National University. This work was supported by NSF grants AGS-2309939, 2401229, and 2408174, and NASA grants 80NSSC24M0174, 80NSSC24K0258, and 80NSSC26K1202. We also acknowledge the computational resources and support provided by Wulver, the high-performance computing cluster at the New Jersey Institute of Technology (NJIT).
\end{acknowledgments}

\bibliographystyle{aasjournal}
\bibliography{
mlsi_references
}

\end{document}